%% file: main.tex
\documentclass[sigconf]{acmart} 
\usepackage{makecell}
\usepackage{algorithm}
\usepackage{algpseudocode}
\usepackage{algorithm,algpseudocode}
\usepackage{hyperref}
\usepackage{xspace}
\usepackage{bm}
\usepackage{gensymb}
\usepackage{subcaption}
\usepackage{float}
\usepackage{listings}
\usepackage{enumitem}
\definecolor{codegreen}{rgb}{0,0.6,0}
\definecolor{codegray}{rgb}{0.5,0.5,0.5}
\definecolor{codepurple}{rgb}{0.58,0,0.82}
\definecolor{backcolour}{rgb}{0.95,0.95,0.95}
\lstdefinestyle{mystyle}{
    backgroundcolor=\color{backcolour},   
    commentstyle=\color{codepurple},
    keywordstyle=\color{NavyBlue},
    numberstyle=\tiny\color{codegray},
    stringstyle=\color{codepurple},
    basicstyle=\ttfamily\tiny,
    breakatwhitespace=true,         
    breaklines=true,                 
    captionpos=t,                    
    keepspaces=false,                 
    numbers=left,                    
    numbersep=5pt,                  
    showspaces=false,                
    showstringspaces=false,
    showtabs=false,                  
    tabsize=2
}

\usepackage{soul}
\usepackage{ulem}
\usepackage{multirow}
\usepackage{makecell}
\input{_defines}

\newcommand{\method}{APPO\xspace}
\newcommand \change[1]{{\textcolor{black}{#1}}}

\definecolor{bcolor1}{rgb}     {1.0,0.0,0.0}
\definecolor{darkgreen}{rgb}     {0.0,0.5,0.0}
\definecolor{blue}{rgb}     {0,0.0,1.0}

\definecolor{red}{rgb}{1, 0, 0}
\definecolor{black}{rgb}{0, 0, 0}
\definecolor{blue}{rgb}{0, 0, 1}

\algnewcommand{\Inputs}[1]{%
  \State \textbf{Inputs:}
  \Statex \hspace*{\algorithmicindent}\parbox[t]{.8\linewidth}{\raggedright #1}
}
\algnewcommand{\Outputs}[1]{%
  \State \textbf{Outputs:}
  \Statex \hspace*{\algorithmicindent}\parbox[t]{.8\linewidth}{\raggedright #1}
}
\algnewcommand{\Initialize}[1]{%
  \State \textbf{Initialize:}
  \Statex \hspace*{\algorithmicindent}\parbox[t]{.8\linewidth}{\raggedright #1}
}

\AtBeginDocument{%
  \providecommand\BibTeX{{%
    \normalfont B\kern-0.5em{\scshape i\kern-0.25em b}\kern-0.8em\TeX}}}

\copyrightyear{2026}
\acmYear{2026}
\setcopyright{cc}
\setcctype{by-nc-nd}
\acmConference[CHI '26]{Proceedings of the 2026 CHI Conference on Human Factors in Computing Systems}{April 13--17, 2026}{Barcelona, Spain}
\acmBooktitle{Proceedings of the 2026 CHI Conference on Human Factors in Computing Systems (CHI '26), April 13--17, 2026, Barcelona, Spain}
\acmPrice{}
\acmDOI{10.1145/3772318.3791443}
\acmISBN{979-8-4007-2278-3/2026/04}

\begin{document}

\title[Preference-Guided Prompt Optimization for Text-to-Image Generation]{Preference-Guided Prompt Optimization for \\ Text-to-Image Generation}

\author{Zhipeng Li}
\affiliation{%
  \institution{Department of Computer Science}
  \institution{ETH Z\"urich}
  \country{Z\"urich, Switzerland}
}
\email{zhipeng.li@inf.ethz.ch}

\author{Yi-Chi Liao}
\affiliation{%
  \institution{Department of Computer Science}
  \institution{ETH Z\"urich}
  \country{Z\"urich, Switzerland}
}
\email{yichi.liao@inf.ethz.ch}

\author{Christian Holz}
\affiliation{%
  \institution{Department of Computer Science}
  \institution{ETH Z\"urich}
  \country{Z\"urich, Switzerland}
}
\email{christian.holz@inf.ethz.ch}

\renewcommand{\shortauthors}{Li et al.}

\begin{abstract}

\input{sections/0_abstract}

\end{abstract}




\begin{teaserfigure}
    \centering%
    \includegraphics[width=\linewidth]{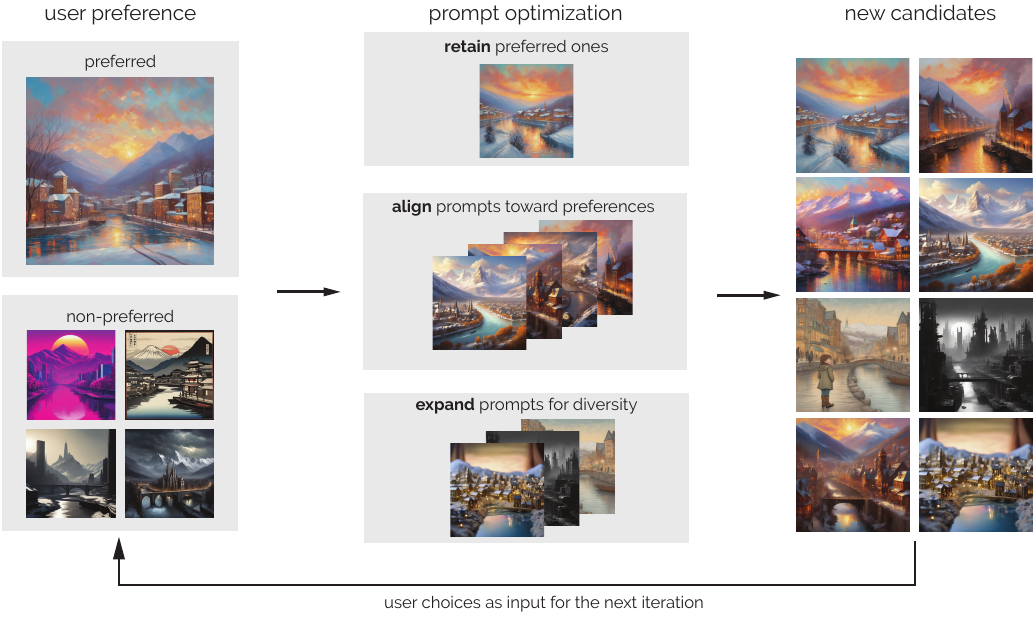}%
    \caption{
        We introduce \method, a prompt optimization framework guided by user preferences. 
        With \method, the user simply needs to iteratively select the generation results they prefer among options that are most aligned with their implicit goals, and APPO will improve the underlying prompts using three strategies: 
        \emph{retain}, which preserves preferred prompts from the previous iteration; 
        \emph{align}, which adjusts non-preferred prompts toward the user's preference; and 
        \emph{expand}, which explores diverse directions by generating new prompts. 
        The outputs generated from the optimized prompts are presented to the user for evaluation in the next iteration. 
        By combining these strategies, our method balances exploration and exploitation, enabling efficient, user-guided prompt optimization with minimal effort.
    }
    \label{fig:teaser}
    \Description{
        This figure illustrates an iterative process for generating images based on user feedback. The workflow begins when a user designates a "preferred" image and several "non-preferred" ones. The system then enters a "prompt optimization" phase where it retains the preferred selection , aligns future prompts toward the user's preferences , and simultaneously expands prompts to create diversity. This results in a set of "new candidates" for the user to evaluate. The process is cyclical, as the user's choices from these new images serve as input for the next iteration, continuously refining the output.
    }
\end{teaserfigure}



\maketitle

\input{sections/1_introduction}

\input{sections/2_related_work}

\input{sections/3_workflow}

\input{sections/4_method}
\input{sections/6_user_study}
\input{sections/7_discussion}

\input{sections/8_conclusion}

\begin{acks}
Yi-Chi Liao was supported by the ETH Zurich Postdoctoral Fellowship Programme.
Zhipeng Li was partially supported by the Swiss National Science Foundation (Grant No. 10004941).
\end{acks}

\balance
\bibliographystyle{ACM-Reference-Format}
\bibliography{sample-base}
\appendix

\clearpage
\input{sections/5_synthetic}

\input{sections/x_appendix.tex}

\end{document}

%% file: _defines.tex
\newcommand{\alphaval}[2]{{\small $p\,#1\,#2$}}
\newcommand{\statsum}[3]{{\small $M=#1\,#3$, $SD=#2\,#3$}}

\newcommand{\friedman}[4]{{\small $\chi^{2}(#1)=#2$, \alphaval{#3}{#4}}}
\newcommand{\wilcoxon}[3]{{\small $Z=#1$, \alphaval{#2}{#3}}}
\newcommand{\ttest}[4]{{\small $t(#1)=#2$, \alphaval{#3}{#4}}}

\newcommand{\anova}[6]{{\small $F_{#1,#2}$\,$=$\,$#3$, $p$\,$#4$\,$#5$}}
\newcommand{\pvall}[2]{{\small $p\,#1\,#2$}} 

%% file: sections/0_abstract.tex
Generative models are increasingly powerful, yet users struggle to guide them through prompts.
The generative process is difficult to control and unpredictable, and user instructions may be ambiguous or under-specified.
Prior prompt refinement tools heavily rely on human effort, while prompt optimization methods focus on numerical functions and are not designed for human-centered generative tasks, where feedback is better expressed as binary preferences and demands convergence within few iterations.
We present \method, a preference-guided prompt optimization algorithm.
Instead of iterating prompts, users only provide binary preferential feedback.
\method adaptively balances its strategies between exploiting user feedback and exploring new directions, yielding effective and efficient optimization.
We evaluate \method on image generation, and the results show \method enables achieving satisfactory outcomes in fewer iterations with lower cognitive load than manual prompt editing.
We anticipate \method will advance human-AI collaboration in generative tasks by leveraging user preferences to guide complex content creation.



%% file: sections/1_introduction.tex
\section{Introduction}

Generative models are becoming increasingly prevalent in everyday life, supporting end-users in various tasks such as writing~\cite{radford2019language, brown2020language}, coding~\cite{fried2022incoder, jiang2024survey}, and image or video generation~\cite{rombach2022high, dhariwal2021diffusion, singer2022make, ho2022imagen}. 
In image generation, where users primarily interact with models through textual prompts that are expected to include all essential information, such as detailed composition of objects, explicit vibes, etc. and often lead to iterative trial-and-error for producing the desired image.
However, crafting effective prompts is challenging~\cite{sahoo2024systematic, DBLP:conf/chi/SubramonyamPPAS24}.
This difficulty arises from the combination of implicit human subjective goals and the black-box nature of generative models.
Users’ generation targets are vague and implicit throughout the generation process~\cite{DBLP:conf/acl/QianHZDQCZZL0024}. 
They usually do not have a precise target in mind; instead, they may only have a general idea of the essential objects and a rough visual style.
They then gradually refine and finalize their goals as they observe intermediate results.
Moreover, even if they have concrete targets in mind, perfectly translating the ideas to effective textual prompts requires experience and trial-and-error.
Simultaneously, the generation process is opaque to users, providing them with little understanding of how the outcomes are exactly generated from the prompts.
Therefore, the prompt creation often relies on iterative trials, which is time-consuming and cognitively demanding, making prompt refinement intrinsically challenging.

Prior works attempted to address this challenge via three rough approaches: supporting users to manually refine the prompt, automatically optimizing prompts without users' involvement, and iteratively refining prompts by incorporating user feedback.
In HCI field, previous studies have explored \emph{various interfaces and tools} to support manual prompt refinement in image generation~\cite{arawjo2024chainforge, brade2023promptify, mishra2023promptaid, feng2023promptmagician}. 
For example, DesignPrompt~\cite{peng2024designprompt} and PromptPaint~\cite{chung2023promptpaint} provide different interfaces that allow users to better combine textual prompts with sketches and color themes to steer the image generation processes. 
PrompTHis~\cite{guo2024prompthis} takes a different approach; it visualizes the outcomes of different prompt variations, allowing the users to visually analyze the effects. 
These works offer interfaces that support users in better editing prompts; however, the decision-making for creating the prompts ultimately relies on humans. 

\emph{Automatic prompt optimization} is an established topic in the machine learning domain. 
Several works focus on training prompt generation models, for instance, by optimizing prompt embeddings with gradient-based methods to maximize a fixed reward~\cite{manas2024improving, dong2023pace, wen2023hard}, exploring prompts that produce more aesthetic outputs via reinforcement learning~\cite{hao2023optimizing}, or dynamically adjusting word weights and timing during generation~\cite{mo2024dynamic}. 
This direction relies on large training data, and the resulting model does not generalize to different tasks well. 
Other works focus on refining prompts using Large Language Models (LLMs) to simulate optimization strategies~\cite{pryzant2023automatic, yuksekgonul2024textgrad} or perform prompt evolution (e.g., edit or mutate existing prompts or combine them)~\cite{fernando2023promptbreeder, guo2025evopromptconnectingllmsevolutionary, cui2024phaseevounifiedincontextprompt, secheresse2025gaapo} to progressively improve performance.
They all operate on numerical evaluation functions to judge the quality of generation results, without a strict search limit.
In reality, this assumption rarely holds: human users cannot reliably and consistently score generation outcomes. 
Rather, humans are better at providing preferences among options~\cite{koyama2020sequential, li2025efficient, DBLP:conf/nips/ChristianoLBMLA17}.
More fundamentally, when a human is involved, it is critical to converge quickly, leading to a stricter budget limit. 

\emph{Iterative prompt refinement that incorporates user feedback} offers a third direction.
DSPy~\cite{khattab2023dspy}, for example, implements a self-improvement LLM pipeline in which prompts are iteratively refined based on users’ natural language feedback. 
ConstitutionMaker~\cite{petridis2024constitutionmaker} transforms user rewrites into structured principles that are appended to the optimizer’s prompt, steering the optimization process toward the user’s preference. 
Similarly, Promptimizer~\cite{wang2025promptimizer} collects user-authored prompt refinements and uses them to guide the selection and ranking of prompts in subsequent iterations.
Although this approach benefits from more explicit and richer information from users, it also imposes a higher cognitive burden, since users must carefully articulate their intents with texts. 
Further, this is not always feasible, as users may not identify the most effective prompts for achieving the optimal results.

Previous work, thus, leaves a clear research gap: 
\textit{How can automatic prompt optimization methods be integrated into human-centered image generation tasks in a way that reduces user effort and remains efficient?} 
Such a method would need to operate from minimal user input—ideally just preference selections rather than manual prompt editing or numerical scoring—while maximizing the utility of this sparse feedback to converge quickly.
There, our goal is to develop a method that supports a novel generative workflow driven \emph{solely by preference feedback} (e.g., selecting from options), thereby minimizing human effort and eliminating the need for manual prompt tuning. 
We further aim to maximize the utility of the sparse feedback from users to ensure high sample efficiency.

In this paper, we introduce Adaptive Preferential Prompt Optimization~(\emph{\method}), a prompt optimization method that is driven by users' preference feedback.
\method begins with a simple initial prompt provided by the user, which should capture the essential elements (e.g., in image generation, specifying the required objects) of the goal, while the details (e.g., adjectives to the objects, and visual style) can be ambiguous or loosely structured, or even missing. 
\method will then iteratively generate a batch of prompts, each leads to a specific generative outcome, and present them to the users. 
The users, instead of carefully fine-tuning the prompts, only need to select the ones that are aligned with their implicit preferences. 
Crucially, users are not required to explain their reasoning, provide evaluation scores, or manually edit the prompts. 
After the selections, \method interprets the human's selection and infers the human preferences, further refining the prompt batch, leading to a set of different outcomes. 
Through repeated preference collection, \method gradually aligns the optimization process with the implicit goals. 
The user's role is limited to simple preference selection, which minimizes effort while still allowing them to meaningfully steer the optimization.

Prompt optimization faces a fundamental challenge: 
searching within the vast space of natural language instructions. 
This challenge becomes even more severe when relying solely on preference selections but not human edits. 
To achieve both effectiveness and sample efficiency, \method must systematically balance exploration, which keeps diversity in prompt proposals, and exploitation, which actively leverages user intent. 
Overemphasizing exploration would require many interactions before convergence, while relying only on exploitation risks premature convergence to local optima.
To address this, \method allocates its limited per-iteration generation budget (i.e., the number of outputs shown to users in each iteration) across three complementary strategies: retainment, alignment, and expansion. 
\textit{Retainment} preserves prompts preferred in the previous iteration, ensuring output can be at least as good as the previous iterations.
\textit{Alignment} exploits user selections by refining non-preferred prompts, shifting them closer in structure or semantics to preferred ones. 
\textit{Expansion} explores the neighborhood around preferred prompts to uncover potentially better solutions.
Finally, to actively regulate exploration and exploitation, we introduce an \emph{adaptation policy} that adapts the expansion range. When the alignment strategy yields semantically similar prompts, indicating convergence toward a local optimum, the policy expands the mutation range to encourage broader exploration. 
Conversely, when alignment already provides sufficiently diverse prompts, expansion restricts its mutation range to support a more focused, efficient search. 
This adaptive mechanism enables \method to steer toward users' preferences while retaining flexibility to escape local optima.

Accompanying our method, we propose a preference-guided generation workflow, which is then evaluated through ablation studies and a user study on image generation tasks. 
In the ablation study, we analyze the contribution of each component by comparing variants without expansion, without alignment, and without adaptive exploration.
We also compared \method with state-of-the-art automatic prompt optimization approaches.
The results show that \method effectively optimizes prompts from binary preference feedback and that its components, together with adaptive exploration, enable the system to escape local optima and converge on better solutions. 
\method also outperformed other prompt optimization methods by achieving better results in fewer iterations and producing better final outcomes.
We then evaluate the workflow with real user prompts in a user study involving two scenarios: 
a target-matching image generation task, where participants reproduced a given image using our workflow versus manually revising prompts, 
and an open-ended image generation task, where participants generated images aligned with their own intent without a reference. 
Results show that our workflow helps users reach satisfying generation outcomes with fewer iterations and less interaction time compared to a state-of-the-art prompt-engineering tool~\cite{wang2024promptcharm} and existing prompt optimization methods~\cite{khattab2023dspy} 
Participants also reported lower cognitive load and less manual effort when using \method, while perceiving a stronger sense of exploration.
Together, these findings demonstrate that our workflow empowers users to control image generation more effectively, achieve desired results with fewer iterations, and experience a more engaging and satisfying creative process.

In summary, we make the following contributions:

\begin{itemize}[leftmargin=*,topsep=3pt]
    \item \method, an automatic prompt optimization approach that leverages users’ binary preference feedback to refine prompts for image generation tasks. 
    It adaptively balances exploiting users' feedback and exploring new directions, enabling users to achieve satisfactory outputs within a few iterations without manual prompt editing.
    \item a preference-driven generation workflow powered by \method. We demonstrate its effectiveness and generalizability through synthetic evaluations and user studies on image generation tasks.
\end{itemize}

%% file: sections/2_related_work.tex
\section{Related Work}

\subsection{Interfaces for Prompt Engineering}

Creating effective prompts remains a non-trivial challenge.  
In HCI domain, prior work has investigated interfaces and tools that support users by visualizing generation results with their prompts, providing refinement suggestions, and enabling multimodal inputs for more precise control.

Since users have limited understanding of how the prompts affect the generation results, previous work has proposed visualizing generation outcomes corresponding to user inputs. 
PrompTHis~\cite{guo2024prompthis} visualized prompts alongside their generated images, helping users understand how specific prompt elements influence outcomes and refine them accordingly. 
PromptMagician~\cite{feng2023promptmagician} introduced interactive visualizations, allowing users to evaluate and filter images based on self-defined criteria and to explore prompt keywords and guidance scales for images of interest. 
PromptCharm~\cite{wang2024promptcharm} assists novice users by mapping parts of the text prompt to regions in the generated image, facilitating more effective refinement. 
For text generation tasks, ChainForge~\cite{arawjo2024chainforge} provided a visual toolkit for improved prompt engineering, while PromptAid~\cite{mishra2023promptaid} allowed users to compare prompt versions and analyze performance differences. 
In 3D generation, Self3DCraft~\cite{xiang2025sel3dcraft} visualized the contribution of each keyword in the prompt alongside the generated 3D mesh.

Beyond visualization, prior work also explored providing prompt suggestions to support refinement. 
Promptify~\cite{brade2023promptify} offers automatic prompt recommendations based on generated image clusters. 
\citeauthor{drosos2025dynamic} proposed generating multiple prompt options from an initial user prompt for selection~\cite{drosos2025dynamic}. 
PromptAid~\cite{mishra2023promptaid} and PromptMagician~\cite{feng2023promptmagician} also provide keyword recommendations alongside visualization.

In addition to text-based prompts, several interfaces support multimodal input to guide generation. 
DesignPrompt~\cite{peng2024designprompt} allows users combining images, colors, and text for more expressive design exploration. 
DirectGPT~\cite{masson2024directgpt} supports users specifying spatial positions and arranging content via drag-and-drop alongside textual prompts. 
PromptPaint~\cite{chung2023promptpaint} enables flexible steering of image generation through interactions resembling traditional painting mediums (e.g., oil, watercolor). 
FusAIn~\cite{peng2025fusain} allows designers to create personalized pens loaded with objects or attributes (e.g., color, texture) to provide additional input during generation.

While these tools facilitate crafting more effective prompts, they still rely on users' judgment and manual edits, which remain cognitively demanding and limit everyday usability for end-users.
Therefore, we aim to optimize the prompts automatically with only users' preference feedback, minimizing users' effort in generation tasks.

\subsection{Prompt Optimization for Image Generation}

There are several challenges in generating images that match a desired outcome using textual prompts. 
First, it is difficult to know in advance the appropriate and enough level of specification needed to express an intended visual quality. 
Further, even when the desired quality is clear, finding the phrasing that produces the intended effect often requires extensive exploration, especially for abstract properties such as visual style. 
These challenges often interact and are mixed. 
\emph{Prompt optimization} is the broad concept that refers to the iteratively improving prompts to achieve a desired outcome. 
\change{
Its goal is to identify the most effective prompt formulation that produces results aligned with the user’s intent. 
Unlike optimizing system-level prompts that govern a model's general behavior to better address a set of tasks~\cite{pryzant2023automatic, yuksekgonul2024textgrad}, the \emph{prompt optimization} concept discussed in this paper focuses on instance-level prompts tailored to generating a specific target output.
This goal is aligned with prior works~\cite{hao2023optimizing, manas2024improving, mo2024dynamic, cui2024phaseevounifiedincontextprompt}.
}
Existing approaches can be grouped into two main strategies. The first relies on refinement of the prompts, in which users iteratively adjust prompts or provide clarifications~\cite{wang2025promptimizer, gao2024aligning}. 
The second is automatic optimization: the system evaluates prompt candidates and infers how to improve output quality without requiring direct edits~\cite{hao2023optimizing, manas2024improving, pryzant2023automatic}.
Below, we review these two categories in turn.

\subsubsection{Refining prompts by human users}
Recent work has explored prompt optimization driven by user textual feedback or rewriting to align the generation results with user preferences and intents.
\citeauthor{gao2024aligning} infer latent user preferences directly from edits: by comparing the original output with user-modified outputs, they learn a latent preference representation, which is then used to condition the model for future generations without requiring numeric scoring~\cite{gao2024aligning}. 
ConstitutionMaker~\cite{petridis2024constitutionmaker} converts natural user rewrites into structured principles that are appended to the prompt context, steering the model to generate outputs consistent with user-specified style, content, or tone. 
Promptimizer~\cite{wang2025promptimizer} collects user-driven prompt refinements and uses them to guide the selection and ranking of prompts in subsequent iterations, effectively aligning outputs to individual preferences. 
DSPy~\cite{khattab2023dspy} is a popular framework for enabling self-improvement in LLMs. 
It can dynamically select an appropriate reasoning strategy (e.g., chain-of-thought~\cite{wei2022chain}, ReAct~\cite{yao2022react}) based on a provided evaluation function or guideline.
As an example use case, DSPy can be used to optimize prompts for image generation: users provide natural-language feedback, and DSPy evaluates multiple prompt variations against the feedback to identify the most effective refinements.

While user textual feedback can provide more information than sparse preference signals, it imposes a higher cognitive load, requiring users to carefully articulate how they want the results to be improved.
In many cases, this is infeasible, as users may not know the precise words to achieve the desired outcome, otherwise they would include them in the initial prompt.
Therefore, we investigate a different approach: prompt optimization in which the system learn from evaluation outcomes and evolve the prompts without human editing.

\vspace{-1em}
\subsubsection{Automatic Prompt optimization with evaluation results}
Recent ML research focused on automatic prompt optimization to improve generation results.  
Most approaches rely on a reward function to evaluate outputs, such as semantic consistency between prompts and generations~\cite{cho2023davidsonian, hu2023tifa, huang2025t2i}, aesthetics for images~\cite{hao2023optimizing, schuhmann2022improved}, or other task-specific metrics.  
With such reward signals, optimization methods can compute gradients and apply learning-based techniques.
Promptist~\cite{hao2023optimizing} fine-tuned a language model with reinforcement learning to transform user inputs into model-preferred prompts. 
PAE~\cite{mo2024dynamic} similarly applied reinforcement learning to adjust word weights and injection time steps, improving generation outcomes.  
DPO-Diff~\cite{wang2024discrete} reduced the search space via LLM-generated synonyms and antonyms, then optimized negative prompts with gradient descent.  
\citeauthor{wen2023hard}~\cite{wen2023hard} proposed mapping prompts into embeddings, applying gradient-based optimization, and decoding back to the nearest natural language expressions.

Beyond embedding-level methods, recent works leverage LLMs directly for prompt-level optimization.  
OPT2I~\cite{manas2024improving} iteratively improves prompts using historical feedback to enhance text-to-image consistency.  
OPRO~\cite{yang2023large} frames optimization as an iterative prompting process, where LLMs are queried with previous histories as context.  
PromptAgent~\cite{wang2023promptagent} treats prompt optimization as a strategic planning problem, applying Monte Carlo tree search to systematically explore expert-level prompts.
Prior research has also explored self-improving prompts using automatically generated checklists~\cite{cook2024ticking, viswanathan2025checklists}. 
These methods derive a checklist from the original instruction and use it to verify whether the generated outputs satisfy each requirement. 
If any checklist item is unmet, the system iteratively refines the prompt until the output fulfills all items.
The growing capabilities of LLMs have also enabled them to simulate classical optimization strategies.  
Automatic Prompt Optimization~\cite{pryzant2023automatic} and TextGrad~\cite{yuksekgonul2024textgrad} introduce natural language “gradients,” drawing analogies to gradient descent for guiding edits.
We also adopt this idea in our implementation to identify effective prompt elements based on user preferences.  
Similarly, evolutionary approaches such as PromptBreeder~\cite{fernando2023promptbreeder}, EvoPrompt~\cite{guo2025evopromptconnectingllmsevolutionary}, PhaseEvo~\cite{cui2024phaseevounifiedincontextprompt}, and GAAPO~\cite{secheresse2025gaapo} use LLMs to generate, mutate, and refine candidate prompts; we also adopt LLM-driven evolutionary search in our implementation for exploration.

While these methods have proven effective, they are typically developed and evaluated using pre-defined numerical reward functions. 
However, they struggle in real-world settings involving human users, as humans cannot reliably or consistently provide numerical ratings for generation outcomes. 
Human ratings are inherently noisy and subjective, due to individual differences in perception and personal preferences.
Consequently, models trained on numerical reward functions are unable to effectively optimize prompts based on actual user scalar feedback.
Although humans are generally better at expressing preferences among options rather than providing absolute ratings, binary preferences convey far less information than scalar ratings. 
Therefore, these models may not converge to a good solution within limited iteration budget by converting these preferences into scalar signals.
Several datasets of human preferences for image generation have been collected and used to fine-tune generative models~\cite{wu2023human, liang2024rich, xu2023imagereward}. 
However, these datasets are typically aggregated across large populations and therefore capture only general human preferences. 
In practice, users’ preferences and generation goals are highly diverse and inherently individual, limiting the applicability of existing preference-based prompt optimization methods.
In contrast, our method aims to directly optimize prompts based on online user preferences in a real-time generation task.

\subsection{Preference-Guided Optimization}

The notion of preference can be formally expressed as a pairwise comparison: an outcome $x$ is preferred over $y$ if $f(x) > f(y)$, where $f(\cdot)$ represents the latent preference function in a user’s mind~\cite{chu2005preference}. 
Compared to providing explicit numerical ratings, it is easier and more reliable for users to indicate which of two options they prefer~\cite{chan2022investigating, yannakakis2015ratings}.
Such pairwise judgments reduce noise and inconsistency in human feedback, making them a valuable signal for optimization. 
Since preferences ultimately guide users’ decisions and behaviors, they serve as an important channel for computational models to better understand users and generate improved outcomes.

A well-established approach to modeling such feedback is to use Gaussian Processes to estimate the probability that $f(x) > f(y)$ from observed comparisons~\cite{eric2007active}. 
By integrating this probabilistic model with Bayesian Optimization, humans can iteratively steer the optimization process through their preferences, an approach known as Preferential Bayesian optimization~\cite{brochu2010bayesian, iwai2025constrained}. 
Building on this line of work, researchers have developed interfaces that make preference elicitation more natural and efficient. 
For instance, \cite{koyama2020sequential} introduced a single slider that aggregates multiple latent factors, allowing users to adjust image appearance without explicitly controlling each dimension; the slider range is refined based on preference feedback. 
Similarly, \cite{koyama2017sequential} proposed a gallery-based interaction where users select from candidate options, and the system adaptively presents the next most promising set. 
This strategy was further improved in efficiency by leveraging prior optimization experience~\cite{li2025efficient}. 
Researchers have also leveraged Gaussian Processes to guide the generation of generative adversarial networks, where the generative space is constrained to specific objects or subjects, and textual prompts are not involved~\cite{nakashima2024swipeganspace}.
In this paper, we adopt the gallery-based paradigm~\cite{koyama2017sequential, li2025efficient} for preference collection in image generation tasks, as it offers a lightweight and intuitive comparison mechanism for users.  

However, when dealing with more complex tasks involving high-dimensional input spaces, traditional probabilistic preference models often struggle to scale.
At the same time, capturing human preference structures becomes increasingly challenging. 
To address this, our work uses preference feedback primarily as a directional signal for optimization, without attempting to explicitly model the entire preference function.

%% file: sections/3_workflow.tex
\section{Preference-Driven Generation Workflow}

Here, we first describe the preference-driven image generation workflow that enables completing generative tasks.
This provides the necessary context for understanding the subsequent preference-guided prompt optimization algorithm and how it operates internally to support the workflow.

\subsection{Prerequisites}
\label{sec:prerequisites}

We define an image generation task as the process in which a user leverages a generative model to produce images for a specific goal. 
The user provides an initial prompt, and our \method edits it iteratively based on the user's preferences until a satisfactory results that meet the specific goal. 
To succeed, prompts must encode two complementary types of information.
The first is \textit{``essential information''}: explicit elements that must appear in the output and are typically clear and easy for users to specify. In image generation, for example, objects such as apple or river represent essential information.
The second is \textit{``implicit intention''}: broader but implicit goals that shape how the essential information should be realized. 
These include aspects such as mood (sad, joyful), visual style (vintage, cold), or object attributes (tall, thin). 
Unlike essential information, implicit intentions are harder to articulate as they may be abstract or ambiguous (e.g., ``a vintage style''), or they may not be a user's immediate focus when drafting an initial prompt.

In our framework, we assume that users can reliably specify the essential information upfront, which anchors the initial prompt. 
This assumption is crucial: the essential information constrains the search space, ensuring that optimization focuses on specific implicit intentions rather than blindly exploring the unbounded prompt space. 
For example, if a user specifies ``a cat sitting on a chair'' as essential information, our method ensures that all generated candidates retain both a cat and a chair. 
The optimization then operates only on reasonable implicit aspects, such as mood (playful or lonely), style (cartoon or photorealistic), or descriptors (wooden chair or metal chair).
Without this constraint, the optimizer might drift toward implicit intentions that are misaligned with the core goal, such as ``surrounded by ocean'' or ``grand medieval battlefield.'' 
These descriptors are semantically rich but typically incompatible with a cat on a chair, leading the system to generate irrelevant or incoherent results. 
We further assume that both essential information and implicit intention remain consistent throughout a single task; if either changes, the process constitutes a new task. 
Under these assumptions, \method focuses on searching the prompt space to discover effective textual representations of the implicit intentions, progressively aligning them with the user's underlying goals.
These assumptions are aligned with the state-of-the-art prompt optimization approaches~\cite{guo2025evopromptconnectingllmsevolutionary, hao2023optimizing, ho2022imagen, lin2024prompt, mahdavi2024ai}.

Below, we provide user interactions within our workflow. 



\begin{figure*}[!h]
    \centering
    \includegraphics[width=\linewidth]{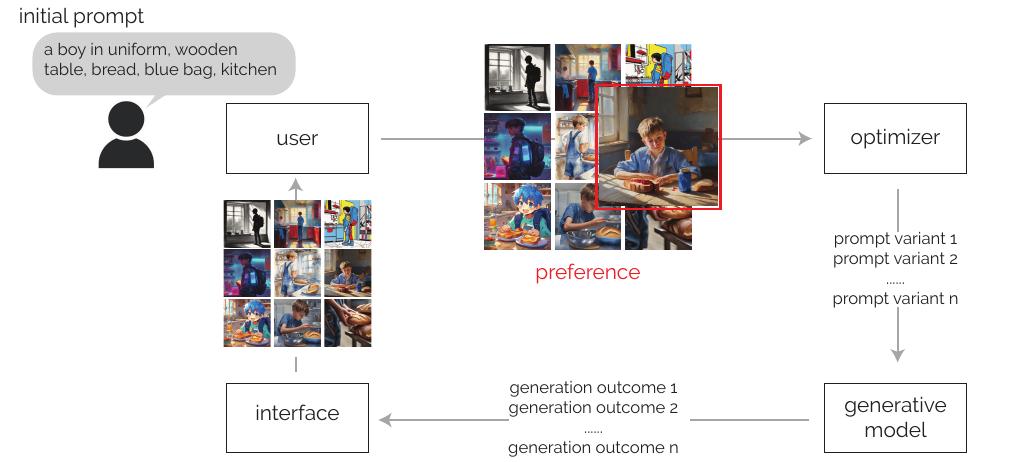}
    \caption{
        Preference-driven generation workflow enabled by \method, with image generation as an example. 
        The user begins with an initial prompt specifying the objects to be included in the generated images (top left). 
        In the first iteration, the optimizer expands this prompt to explore diverse possibilities and generate multiple prompt variants (right). 
        The generative model then produces outputs corresponding to these variants (down), which are presented to the user in a gallery (left). 
        The user selects their preferred results (up), which are then fed into the optimizer who infers their preferences to generate refined prompts. 
        These refined prompts are used by the generative model to produce new outputs, which are further evaluated by the user in subsequent iterations.
    }
    \label{fig:workflow}
    \Description{
    This figure illustrates an iterative AI image generation workflow where a "user" provides an initial text prompt and selects a "preference" from a set of generated images. An "optimizer" uses this preference to create multiple new "prompt variant n" , which are then used by a "generative model" to produce new images. These "generation outcome n" are displayed back to the user via an "interface", allowing the user to make a new selection and repeat the refinement cycle.
    }
\end{figure*}

\subsection{Workflow}
\label{sec:workflow}

The workflow starts with the \emph{user} providing an initial prompt, as illustrated in~\autoref{fig:workflow}, using image generation as an example.  
This prompt serves as a starting point: it should include essential elements of the task (e.g., specifying core objects in image generation) but may remain underspecified or vague with respect to style, composition, or other modifiers.
In the first iteration, the \emph{optimizer} expands the initial prompt in two ways: it completes missing details to form viable prompts, and it explores diverse directions in the prompt space.
Since no preference information is available yet, the focus at this stage is to explore a broad range of possibilities.
The expanded prompts are then passed to the \emph{generative model}, yielding a set of candidate outputs.
The candidate outputs are displayed to the \emph{user} through an \emph{interface}, where the user selects one or more results that best capture their implicit goal. 
This feedback is then passed to the \emph{optimizer}, which analyzes differences between preferred and non-preferred prompts to infer user preferences and underlying intent. 
Based on these insights, the \emph{optimizer} generates refined prompt candidates that place greater emphasis on directions aligned with the user’s evolving goals. 
The \emph{generative model} transforms these prompts into a new set of outputs for \emph{user} to evaluate. 
Through this iterative cycle of selection, analysis, and refinement, the prompts progressively converge toward results that satisfy the user’s intent.

\subsection{Core elements}

Our preference-driven generation workflow contains four interconnected core elements: the human user, the interface for presenting results, the prompt optimizer, and the generative model.
Together, these components form a closed feedback loop through which implicit user goals are progressively externalized and the outputs are increasingly aligned with those goals.

\emph{User.}
The human user initiates the generation task by providing an initial prompt and subsequently evaluating generated outputs at each iteration.
Rather than editing prompts iteratively, users simply select the results that best align with their implicit goals.
This interaction is designed to be lightweight: preferences are expressed in a binary manner (selected vs. not selected), reducing both cognitive and manual effort while still providing meaningful feedback to guide optimization.

\emph{Interface.}
Generated results are presented to the user through an interface designed to support efficient inspection and preference expression. 
In image generation, outputs are typically arranged in a gallery, as in prior work~\cite{koyama2020sequential, li2025efficient} and illustrated in~\autoref{fig:workflow}, allowing users to visually compare multiple candidates at once. 
While galleries are effective for visual media such as images or videos, alternative layouts may be more appropriate in other domains to allow users to access outputs and provide binary feedback.
For example, text-based tasks may benefit from presenting fewer candidates at a time for more detailed reading. 

\emph{Optimizer.}  
The optimizer refines prompts using \method in each iteration. 
It takes the user's preference feedback and the current set of prompt candidates as input, while also maintaining a history of prompts explored in previous iterations. 
Based on this information, it outputs a new set of optimized prompts for the generative model. 
We employ an LLM to perform the operations in the optimizer, ensuring that the optimized prompts remain human-readable. 
This preserves transparency and allows for optional user intervention when desired. 
In contrast, embedding-based approaches often sacrifice interpretability, making them less suitable for workflows in which users actively participate.
We describe the optimizer's internal mechanisms in~\autoref{sec:method}.

\emph{Generative Model.}  
The generative model translates refined prompts into concrete outputs. 
Depending on the task, this may involve LLMs for text generation, diffusion models for image synthesis, motion generation models for motion creation, or other domain-specific generators. 
As the realization layer of the workflow, the generative model produces the artifacts that are presented to users for evaluation, thereby closing the feedback loop that drives subsequent iteration.

%% file: sections/4_method.tex
\section{Adaptive Preferential Prompt Optimization}
\label{sec:method}

In this section, we introduce Adaptive Preferential Prompt Optimization (\method), our adaptive prompt optimization method that refines prompts based on user preferences. 
We begin by outlining the key challenges of preference-guided prompt optimization will face, then present the working principles of our method, and finally provide a detailed description.

\subsection{Challenges}

As described in the workflow, preference-guided prompt optimization must rely on users' feedback while aiming to fulfill their implicit generation goals efficiently.
Meeting these requirements gives rise to several challenges, which we detail below.

\textbf{Challenge 1: How to leverage sparse preference signals to guide prompt optimization?}
In generation tasks, the prompt search space is enormous, yet users can only evaluate a small subset of prompts because human evaluation is both time-consuming and cognitively demanding. 
Consequently, user feedback is inherently sparse, covering only a tiny fraction of possible prompts. 
Moreover, users often struggle to provide consistent numerical ratings; instead, binary preference feedback (e.g., indicating which option is preferred) has been shown to provide a more accurate and reliable signal~\cite{koyama2017sequential, koyama2020sequential, li2025efficient}. 
The key challenge is, therefore, how to maximize the utility of these sparse preference signals to effectively guide the next iteration of prompt generation.  

\textbf{Challenge 2: How to maintain diversity and escape local optima in prompt optimization?}
The opposite problem arises when the optimizer relies too heavily on exploiting user preferences. 
Overfitting to a small set of selections can cause all prompts to converge too quickly toward a narrow region of the search space. 
This reduces the diversity of subsequent generations, forcing users to review iterations of highly similar outcomes and preventing the discovery of their broader options. 
In optimization terms, the process is known as being trapped in a local optimum. 
How to maintain diversity and systematically escape premature convergence is therefore another key challenge.


\textbf{Challenge 3: How to strike the balance between exploration and exploitation for efficient prompt optimization?}
Building on the prior two challenges, a more fundamental issue is how to balance exploitation and exploration.  
Exploitation leverages user preferences to refine prompts in a stable direction but risks reducing diversity.  
Exploration, in contrast, probes new regions of the prompt space, which may uncover better solutions but also risks producing inferior results and increasing instability.  
Striking the right balance between these two strategies is a known dilemma in general optimization, but it becomes especially critical in human-involved tasks where evaluations are costly.  
Here, sample efficiency is an important constraint: the optimizer must approach high-quality prompts in as few iterations as possible while still preserving the ability to escape local optima.

\subsection{Prompt Generation Strategies}

\label{sec:prompt_strategies}
In each iteration of the workflow, the system generates $n$ outcomes based on prompts, presents them to the user, and then receives binary preference feedback.  
Here, $n$ denotes the per-iteration \textit{generation budget}. 
This budget is divided across complementary strategies, which are designed to address the challenges outlined above.


\textbf{Strategy 1: Retain user-preferred prompts (Retainment).}  
A straightforward way to exploit preference signals observed thus far (addressing Challenge~1) is to retain the most preferred prompts from previous iterations.  
This strategy, commonly used in preferential optimization methods~\cite{koyama2017sequential}, provides stability by ensuring that output quality either improves (when users prefer a newly generated outcome) or remains consistent (when they re-select the same option).  
From the user's perspective, this creates a sense of steady progress, reinforcing trust in the system and supporting a positive overall experience.  


\textbf{Strategy 2: Align prompts toward the preferred selections (Alignment):}  
Instead of simply discarding non-preferred prompts, the optimizer leverages them by comparing them against preferred ones (Challenge~1).  
This comparison reveals which qualities or aspects better align with the user's implicit preferences.  
The resulting differences can be interpreted as a ``text gradient'' (e.g., \citet{pryzant2023automatic}), which is then applied to adjust non-preferred prompts, moving them closer to the user's intent.  
Implementation details of how this gradient is estimated and applied are provided later.


\begin{figure*}[!t]
    \centering
    \includegraphics[width=\linewidth]{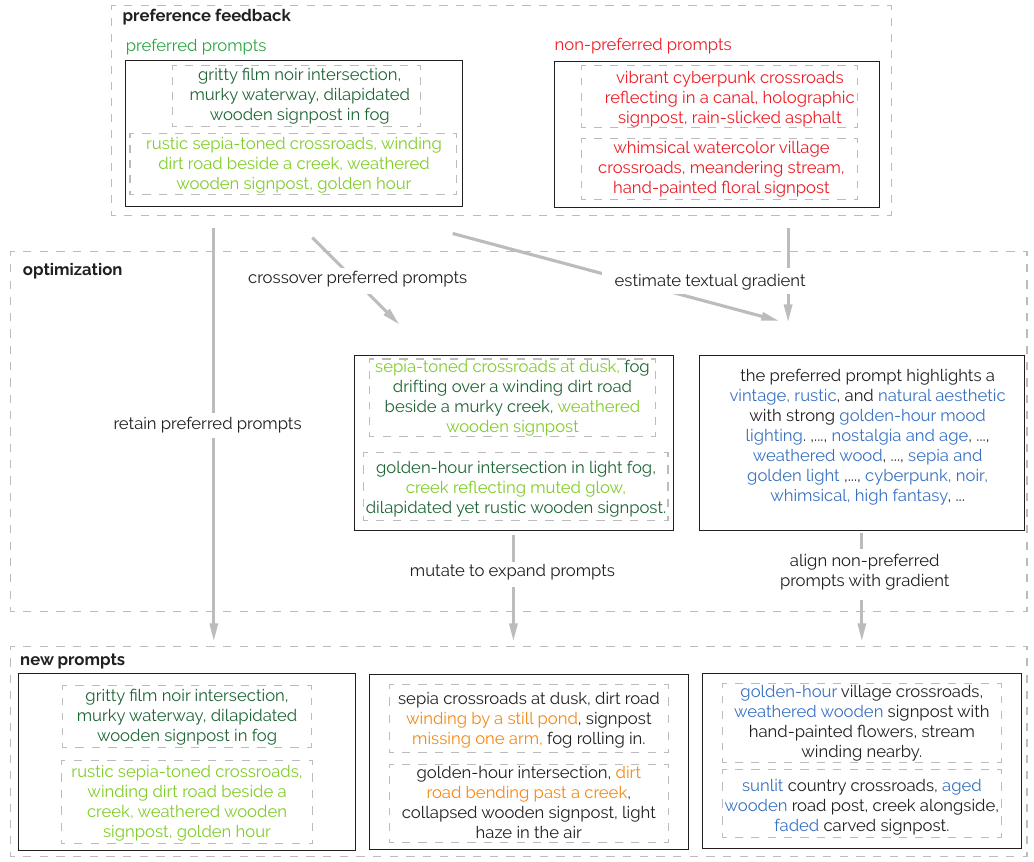}
    \caption{The concept of optimization methods behind \method with a concrete example.
        In each iteration, \method takes the preferred prompts (green) and non-preferred prompts (red) from the previous iteration as input. 
        Three strategies are applied: It first \emph{retains} the preferred prompts.
        In parallel, the \emph{expansion} strategy applies evolutionary operations to explore new prompts. 
        This begins with \emph{crossover}, mixing elements of the preferred prompts (words in different green colors), followed by \emph{mutation} to generate additional prompts (orange).
        Simultaneously, the \emph{alignment} strategy estimates the textual gradient from both preferred and non-preferred prompts to identify elements generally favored by the user (blue).
        This gradient is then applied to non-preferred prompts to better align them with user preferences.
        Finally, all of the three \emph{retainment}, \emph{expansion}, and \emph{alignment} prompts form the set of prompts for the next iteration.
    }
    \label{fig:optimization_workflow}
    \Description{
    This figure details a prompt "optimization" workflow that starts with user "preference feedback" , where specific prompts are labeled as "preferred" , such as "rustic sepia-toned crossroads" , and others as "non-preferred" , like "vibrant cyberpunk crossroads". The system then "estimate[s a] textual gradient" to identify the user's preferred aesthetic, which includes a "vintage, rustic, and natural aesthetic" with "sepia and golden light". This analysis drives four simultaneous actions: the system will "retain preferred prompts" , "crossover preferred prompts" to create hybrids like a "sepia-toned crossroads at dusk" with fog , "mutate to expand prompts" , and "align non-preferred prompts with gradient". This multi-step process results in a list of "new prompts" that includes the originals , new combinations , and aligned versions of previously rejected concepts.
    }
\end{figure*}

\textbf{Strategy 3: Expand the prompt search space for diversity (Expansion).}  
Continued exploration of the prompt space is essential to avoid local optima and to encourage discovery of novel solutions (Challenge~2).  
A common approach is to vary prompts using evolutionary principles, including operations such as adding, deleting, mutating, or recombining words (e.g., \citet{guo2025evopromptconnectingllmsevolutionary}). 
The degree of change directly determines the search range: 
Unconstrained mutations produce a broad but largely random exploration, and guided edits on specific words focus the search on a narrow region and lean toward exploitation behavior.  
This trade-off (Challenge~3) is essentially to balance exploration and exploitation. 

To address this, we introduce a \textbf{policy for controlling exploration intensity}.
At the high level, the policy ensures the overall optimization moves from exploration to exploitation across iterations; that is, the diversity within the prompts in each iteration should gradually decrease. 
Beyond that, the policy also ensures enough exploration within each iteration, avoiding the mutation also highly focused on the selected prompts from the previous rounds. 
Specifically, \method computes the semantic similarity between retained prompts and newly aligned prompts to assess whether the aligned prompts explore diverse regions.
If they are highly similar, the policy encourages random, less-constrained exploration; otherwise, it restricts exploration to the local vicinity of promising regions. 
Overall, this policy adapts the expansion strategy, making the prompt generation controlled and efficiently balancing diversity with convergence.

\subsection{Implementation of APPO}

\begin{algorithm*}[t]
\caption{Preference-Guided Prompt Optimization}
\label{alg:full_pipeline_feedback}
\begin{algorithmic}[1]
\Require 
$P_0$: initial prompt, 
$T$: maximum iterations, 
$G(\cdot)$: generative model, 
$n$: per-iteration generation budget
\State Initialize iteration counter: $t \gets 1$
\State Initialize prompt set: $P \gets P_0$
\State Initialize history: $H_p \gets \emptyset, \; H_n \gets \emptyset$
\State Initialize exploration parameters: $v \gets 0, \; v_{\min} \gets 0, \; v_{\max} \gets 0$
\While{$t \leq T$}
    \If{$t = 1$} 
        \State $P' \gets \mathrm{Randomize}(P)$ 
        \Comment{Initial exploration for diversity}
    \Else
        \State Retainment: $R \gets P_p$ 
        \Comment{Keep user-preferred prompts}
        \State Alignment: $A \gets \mathrm{Align}\!\left(P_p, P_n, H_p, H_n, \left\lfloor \tfrac{n - |R|}{2} \right\rfloor \right)$
        \Comment{Balance retained and non-preferred prompts}
        \State Update intensity: $(v, v_{\min}, v_{\max}) \gets \mathrm{UpdateIntensity}(R, A, v_{\min}, v_{\max})$
        \Comment{Adjust exploration strength}
        \State Expansion: $E \gets \mathrm{Expand}(P_p, \; n - |R| - |A|, \; v)$ 
        \Comment{Generate variations around retained prompts}
        \State Candidate set: $P' \gets R \cup A \cup E$
        \State Consistency Check: $P'' \gets \mathrm{Reflect}(P_0, P')$ 
        \Comment{Ensure key information from $P_0$ is preserved}
    \EndIf
    \State Generate outcomes: $C \gets G(P'')$
    \State Collect feedback: $(P_p, P_n) \gets \mathrm{UserFeedback}(C)$ 
    \Comment{$P_p$: preferred, $P_n$: non-preferred prompts}
    \State Update history: $H_p \gets H_p \cup P_p, \; H_n \gets H_n \cup P_n$
    \If{user is satisfied with any $P_p$}
        \State \Return final optimized prompt $P_p$
    \EndIf
    \State $P \gets P'$
    \State $t \gets t + 1$
\EndWhile
\State \Return final optimized prompt $P_p$ and the satisfied generation outcomes
\end{algorithmic}
\end{algorithm*}

With the working principles identified above, we implemented our preference-guided prompt optimization method as summarized in~\autoref{alg:full_pipeline_feedback}.

\subsubsection{Overview}
Our method starts with an initial prompt $P_0$ and refines it continuously based on user feedback until the user is satisfied or a predefined maximum number of iterations is reached. 
Importantly, the initial prompt should contain the essential information that must be correctly presented, such as objects to include. This assumption and our intended optimization goal are introduced in~\autoref{sec:prerequisites}.
In the first iteration, before any preference feedback is given, the optimization performs random exploration to generate a diverse set of candidate prompts from the initial input. 
From the second iteration onward, new prompts are generated through three strategies: \textit{retainment, alignment}, and \textit{expansion} (see \autoref{sec:prompt_strategies}).
Specifically, the generation budget $n$ first reserves slots for the items selected in the previous iteration, and then divides the remaining budget equally between alignment and expansion.
A \emph{consistency check} step follows to ensure that essential information from the initial prompt is preserved. 
The generated candidates are then passed to the generative model, and the resulting outputs are presented to the user, who selects preferred ($P_p$) and non-preferred ($P_n$) options. 
This feedback informs the next iteration of prompt generation. In the following sections, we provide a detailed description of each step of \method.



\subsubsection{Random Exploration}
In the first iteration, no historical information about the user's preferences is available.
The optimizer only has access to the initial prompt, so it can only rely on random exploration to generate diverse candidate prompts. 
We leverage an LLM to rewrite the initial prompt into complete and well-formed variants.
Each variant explores different directions (e.g., in image generation, diverse styles, visual details, or wording) while preserving the core objects and descriptors of the original prompt.
This approach ensures that the generated candidates are both coherent and highly diverse, providing a rich starting point for subsequent iterations.


\subsubsection{Retainment}
In each iteration, \method always applied one strategy, which directly carries over the prompts marked as preferred by the user in the previous round. 
These retained prompts are included unchanged in the new candidate pool, ensuring that best options remain available for the next iteration.

\subsubsection{Alignment}
The second strategy involves aligning non-preferred prompts toward the preferred ones. 
We achieve this using a textual gradient approach~\cite{pryzant2023automatic}, inspired by gradient descent in continuous optimization. 
In numerical optimization, gradients provide a direction for improving performance.  
In prompt optimization, however, true gradients are unavailable, since the only way to assess the quality of prompts is feeding it to the target model, generating the outcome, followed by user evaluation.
Prior research~\cite{pryzant2023automatic, yuksekgonul2024textgrad} has demonstrated that LLMs can approximate a ``textual gradient'' by comparing different prompts. 
The model analyzes differences between these prompts to infer which components contribute positively to desired outcomes and which should be modified or removed. 
This information acts as a directional signal: it guides the transformation of non-preferred prompts toward patterns found in the preferred prompts, effectively “nudging” them in the direction of user-aligned language without requiring explicit human rewriting.


Given the sparsity of human feedback, our method incorporates the history of prompts from previous iterations when estimating the textual gradient. 
By summarizing the history of preferred and non-preferred prompts, the LLM can better identify recurring effective elements than only relying on the current selections.
Using this information, the non-preferred prompts are gradually adjusted toward the preferred prompts while keeping their other characteristics, allowing the optimizer to exploit user-indicated elements efficiently without requiring extensive evaluation.

The procedure is formalized in~\autoref{alg:text_gradient}.
First, our method summarizes ($\mathrm{LLM}_{s}$) the history of preferred ($H_p$) and non-preferred prompts ($H_n$) to capture the effective elements that consistently lead to desirable outcomes ($S$). 
Next, it computes a textual gradient ($g$) between the preferred ($P_p$) and non-preferred prompts ($P_n$), highlighting the elements to retain and the elements to modify. 
Finally, our method uses this gradient, together with the summarized history, to iteratively generate (using $\mathrm{LLM}_{\delta}$) a set of new prompts that are closely aligned with the user’s preferences, ensuring both high quality and diversity in the candidate set for the next iteration.

\begin{algorithm}[!h]
\caption{$\mathrm{Align}(\cdot)$, corresponding to line 10 in \autoref{alg:full_pipeline_feedback}}
\label{alg:text_gradient}
\begin{algorithmic}[1]
\Require 
$P_p$: preferred prompts from the last iteration; 
$P_n$: non-preferred prompts from the last iteration
$H_p$: history of preferred prompts; 
$H_n$: history of non-preferred prompts; 
$m$: number of gradient-based candidates
\Return 
$P'$: aligned candidate prompts for the next iteration

\State $P' \gets \emptyset$
\State $S \gets \mathrm{LLM}_{s}(H_p, H_n)$ 
\Comment{Summarize historical prompts to extract salient patterns}
\State $g \gets \mathrm{LLM}_{\nabla}(P_p, P_n)$ 
\Comment{Estimate textual gradient from preferred vs.\ non-preferred prompts}

\For{$i \gets 1$ to $m$}
    \State $p'_i \gets \mathrm{LLM}_{\delta}(S, g)$ 
    \Comment{Generate a new candidate aligned with preferred features}
    \State $P' \gets P' \cup \{p'_i\}$
\EndFor

\State \Return $P'$
\end{algorithmic}
\end{algorithm}

\subsubsection{Expansion}

The third operation further explores candidate prompts based on the preferred prompts.
Because prompts are discrete natural language, we leverage LLMs to perform an evolutionary algorithm (EA) to evolve the textual prompts. 
Prior work has demonstrated that this approach is effective for evolving prompts~\cite{fernando2023promptbreeder, guo2025evopromptconnectingllmsevolutionary}.

Such LLM-based EA works as follows. 
First, a subset of the current candidate prompts is selected to serve as \emph{parents}. 
These parents form the foundation for generating new prompts. 
Next, a \emph{crossover} operation combines elements from multiple parents, such as tokens, phrases, or structural patterns, to produce new child prompts that inherit characteristics from more than one high-quality candidate. 
Following this, a \emph{mutation} step introduces additional variations, such as replacing or reordering words, inserting modifiers, or adjusting sentence structure. This step ensures diversity in the candidate pool and helps the algorithm escape local optima.
Finally, a \emph{selection} step evaluates all generated children according to a fitness function. 
Only high-quality candidates are retained for the next iteration. 
By repeatedly applying this cycle of selection, crossover, and mutation, the algorithm progressively expands the search around promising regions, improving the chances of discovering prompts that better align with the user's intent.

When adopting EAs in our method, human users provide the preferred prompts, which serve as parent prompts. 
It then generates new candidates through a crossover process that combines features from multiple parents. 
Each resulting child prompt is then mutated by the LLM to introduce additional variation, controlled by the mutation number and intensity. 
Finally, we select a subset of children using a fitness function.
The fitness function evaluates each mutated child prompt based on two complementary objectives: similarity to the parent prompts and diversity among the children. 
Similarity ensures that the generated prompts remain aligned with the user’s preferred directions, while diversity encourages exploration of the local search space. 
We compute a Pareto front over these two objectives (similarity and diversity) to identify prompts that optimally trade off alignment and variation. 
From this Pareto front, a fixed number of children are selected for the next iteration, guaranteeing that the resulting candidate set is both high-quality and diverse, efficiently guiding the optimization process.
This iterative process allows the system to efficiently explore the local neighborhood around preferred prompts while maintaining alignment with user preferences.
The procedure is formalized in~\autoref{alg:ea}.

\begin{algorithm}[t]
\caption{$\mathrm{Expand}(\cdot)$, corresponding to line 12 in \autoref{alg:full_pipeline_feedback}}
\label{alg:ea}
\begin{algorithmic}[1]
\Require 
$P_p$: preferred prompts from the last iteration;  
$k$: number of children to select;  
$v$: mutation intensity  
\Return 
$S$: selected candidate prompts for the next iteration  

\State $C \gets \mathrm{LLM}_{c}(P_p)$ 
\Comment{Crossover: generate children from parent prompts}
\State $M \gets \emptyset$
\For{$i \gets 1$ to $|C|$}
    \State $m_i \gets \mathrm{LLM}_{m}(C_i, v)$ 
    \Comment{Mutate each child with adaptive intensity $v$}
    \State $M \gets M \cup \{m_i\}$
\EndFor

\State $F \gets \emptyset$ 
\Comment{Evaluate fitness of mutated prompts}
\For{$i \gets 1$ to $|M|$}
    \State $s_i \gets \mathrm{Sim}(m_i, P_p)$ 
    \Comment{Similarity to parent prompts}
    \State $d_i \gets \mathrm{Div}(m_i, M \setminus \{m_i\})$ 
    \Comment{Diversity relative to other children}
    \State $F_i \gets (s_i, d_i)$
    \State $F \gets F \cup \{F_i\}$
\EndFor

\State $S \gets \mathrm{SelectPareto}(M, F, k)$ 
\Comment{Select $k$ Pareto-optimal children balancing similarity and diversity}
\State \Return $S$
\end{algorithmic}
\end{algorithm}

\subsubsection{Policy for Adaptive Expansion}

While the three strategies of generating new prompts ensure coverage of prompts close to the user-preferred ones and local exploration, the system may still get stuck in local optima. 
In particular, the only mechanism for exploration is the mutation step in the EAs. 
When all parent prompts are highly similar, a default mutation that only makes minor word changes can fail to sufficiently explore the broader prompt space.

To address this, we incorporate a \emph{policy for adaptive control of mutation intensity}. 
The default mutation modifies only a few words without changing the overall meaning or composition. 
However, LLMs can rewrite prompts in more substantial ways, allowing for both subtle and extensive exploration. 
We introduce a mutation intensity variable $v \in [0,1]$ to control the mutation behavior: 
When $v=0$, the mutation is minimal, and only small word-level changes without altering the core meaning.
On the contrary, when $v=1$, mutation is maximal, and the optimizer performs random exploration as in the first iteration.

In our method, the mutation intensity is determined based on the semantic similarity between the Retained prompts (preferred prompts from the last iteration) and the Aligned prompts (non-preferred prompts aligned to preferred ones). 
If the Aligned prompts are very similar to the Retained prompts, the system increases the mutation intensity to explore a wider area. 
Conversely, if the similarity is low, indicating that the Aligned prompts already introduce local variations, the mutation intensity remains low to avoid drifting too far from user-preferred prompts.

To quantify semantic similarity between prompts, we employ the CLIP model~\cite{DBLP:conf/icml/RadfordKHRGASAM21} to encode prompts into embeddings within a shared representation space.
We then compute cosine similarity between the embedding vectors, where higher values indicate greater semantic similarity (i.e., a smaller angular difference between the vectors).
This similarity serves as the basis for controlling mutation intensity during optimization.
A naive approach would be to directly map the absolute similarity values to mutation intensity.  
However, absolute similarity scores are difficult to interpret: the same value can correspond to very different levels of semantic relatedness depending on the task and context.  
For example, in an image generation task, the prompts ``a red apple'' and ``a green apple'' may yield an extremely high similarity score, despite producing visually distinct outputs.  
Conversely, the prompts ``a magical forest'' and ``a fantasy-world woodland'' may produce a lower similarity score, even though the resulting outputs are expected to be nearly indistinguishable.  
Thus, relying on absolute similarity values would make mutation intensity overly sensitive to task context rather than reflecting meaningful differences between prompts.  
To address this, we normalize similarity scores dynamically for each task: in each iteration, the similarity value is rescaled relative to the observed upper and lower bounds from previous iterations.
This normalization produces values that more reliably reflect the relative diversity of prompts, independent of the absolute scale of the task.
The resulting normalized similarity determines the mutation intensity for the current iteration, after which the bounds are updated.
Adaptive expansion begins from the third iteration, once sufficient data are available to initialize the bounds.

This adaptive strategy ensures a dynamic balance between local exploitation and broader exploration, allowing the prompt optimization process to escape local optima while respecting user preferences.

\begin{algorithm}[t]
\caption{$\mathrm{UpdateIntensity}(\cdot)$, corresponding to line 11 in \autoref{alg:full_pipeline_feedback}}
\label{alg:adaptive_mutation}
\begin{algorithmic}[1]
\Require 
$R$: retained prompts;  
$A$: aligned prompts;  
$v_{\min}, v_{\max}$: historical similarity bounds  
\Ensure 
$v$: updated mutation intensity;  
$v_{\min}, v_{\max}$: updated similarity bounds  

\State Compute pairwise similarities: $s_{ij} \gets \mathrm{Sim}(R[i], A[j])$
\State $s_{\text{avg}} \gets \mathrm{mean}(s_i)$
\State Normalize similarity: 
$v \gets \dfrac{v_{\max} - s_{\text{avg}}}{v_{\max} - v_{\min}}$
\State Clamp $v \gets \min(\max(v, 0), 1)$

\State Update bounds: 
$v_{\min} \gets \min(v_{\min}, s_{\text{avg}})$;  
$v_{\max} \gets \max(v_{\max}, s_{\text{avg}})$
\State \Return $(v, v_{\min}, v_{\max})$
\end{algorithmic}
\end{algorithm}

\subsubsection{Consistency Check}


At the final stage of optimization, we introduce a consistency check step that ensures all essential information from the initial prompt is preserved in the optimized outputs.
Although the alignment and expansion strategies refine the prompt, they may also omit or distort crucial elements from the initial prompt.
For example, in image generation tasks, users typically specify the objects that should appear in the initial prompt, yet these objects may be removed or altered during optimization.
To prevent such issues, we employ a separate LLM instance to perform a structured comparison between the original and optimized prompts.
This comparison extracts the essential elements from the initial prompt (e.g., objects to include and their descriptors), checks whether they are preserved, and restores any that are missing.
Through this process, the consistency check step ensures that prompt optimization progresses toward user preferences while aligning with the user's original, core intent.

\subsection{Synthetic Tests}

We evaluated \method through synthetic tests to assess its effectiveness and to evaluate the contributions of our optimization strategies and policy (see Supplementary Material). 
In particular, we examined variants of \method that omit key components: expansion, alignment, or adaptive exploration, in order to analyze the contribution of each.
To simulate generation goals, we modified initial prompts from existing datasets by adding a specific visual style. 
User preferences were simulated by computing the similarity between the generation results of candidate prompts and those of the target prompt.

Our results show that variants without expansion or without alignment stuck at suboptimal solutions after approximately 6–8 iterations, yielding only a limited improvement (9.38\% for without expansion and 9.87\% for without alignment) compared to simply paraphrasing the initial prompt. 
The variant without adaptive exploration achieves a larger improvement (14.66\%), but fails to make further progress after around 10 iterations. 
In contrast, \method reaches a better solution within approximately 5 iterations and continues to improve, ultimately achieving an improvement of 19.64\%. 
These findings demonstrate that \method can effectively optimize prompts based on sparse feedback, while highlighting the contribution of each component.

We further compared \method to two state-of-the-art automatic prompt optimization methods, TextGrad~\cite{yuksekgonul2024textgrad} and Self-TICK~\cite{cook2024ticking}, both of which aim to align generation results with prior observations.
The results show that \method consistently outperforms both baselines. 
While the baselines reach their best performance (6.45\% for TextGrad and 8.06\% for Self-TICK) after approximately five iterations, they fail to make further improvements.
These findings suggest that \method is more effective than existing automatic prompt optimization methods for image generation tasks, even when relying solely on preference-based feedback.


%% file: sections/6_user_study.tex
\begin{figure*}[!ht]
    \centering
    \includegraphics[width=\linewidth]{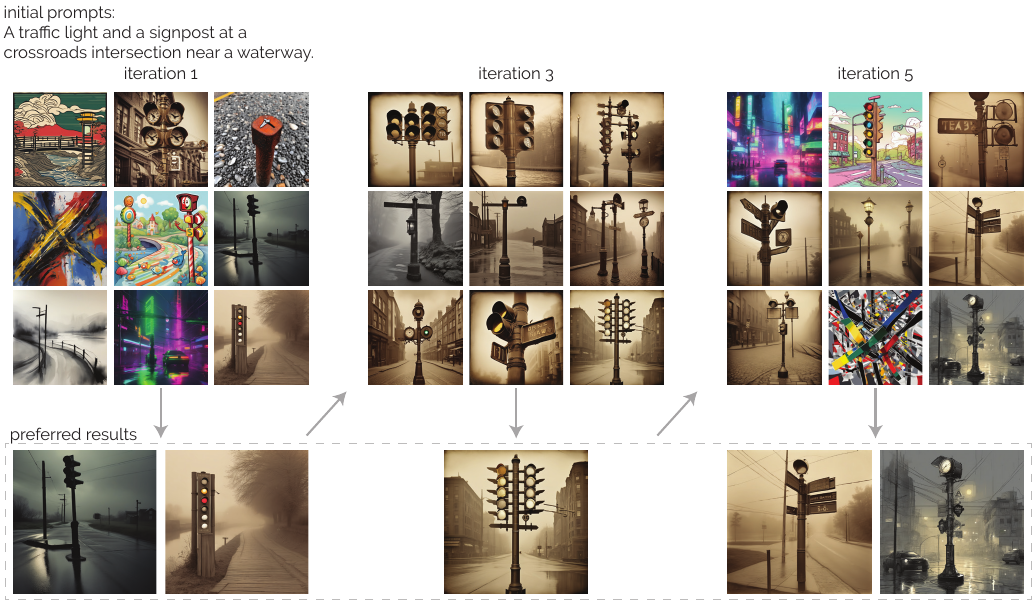}
    \caption{User study interface and intermediate results from \method. 
    Participants are presented with nine candidate images and select one or more that best align with their generation goal. 
    They can indicate whether they are satisfied with the current iteration’s outcome or wish to see additional candidates. 
    In the results shown here, iteration 3 (middle) converges toward the “old-film” theme selected by the user in previous iterations (left), demonstrating that \method effectively guides generative models toward user preferences. 
    By iteration 5 (right), \method detects that the optimization has converged to a local optimum and responds by exploring unknown directions to continue improving the results.
    }
    \label{fig:study_platform}
    \Description{
    This image displays an AI image generation process across "iteration 1" , "iteration 3" , and "iteration 5" , based on the initial prompts "A traffic light and a signpost at a crossroads intersection near a waterway". The initial set of images is stylistically diverse, but as the user repeatedly selects their "preferred results"—which are consistently dark, moody, and atmospheric scenes of traffic lights in fog or rain—the subsequent generations of images progressively converge on this specific vintage and melancholic aesthetic.
    }
\end{figure*}

\section{User Study}

Alongside the synthetic tests, we conducted a user study comparing \method with three representative alternatives: 
a state-of-the-art prompt-engineering tool for image generation~\cite{wang2024promptcharm}, 
a prompt-optimization method that incorporates user feedback~\cite{khattab2023dspy}, 
and a prompting method that helps users to specify their goals in their prompts by asking them clarification questions~\cite{hahn2024proactive}.


\subsection{Design}

To investigate if our method could help users to obtain generation outcomes that aligned with their intent with less effort, we adopted a within-user study design including four conditions:


\begin{itemize}
    \item \emph{PromptCharm}:
        Participants used PromptCharm~\cite{wang2024promptcharm}, a state-of-the-art prompt engineering interface for image generation. 
        PromptCharm suggested automated refinements to users’ prompts, allowed users to adjust the attention given to each word, provided image style suggestions, and supported comparison of generation results across iterations. 
        It also helped users understand the correspondence between prompt words and parts of the generated image, further aiding in prompt refinement. We directly employed the original code and interface for our study \footnote{\url{https://github.com/ma-labo/PromptCharm}}.
    \item \emph{DSPy}:
        In this condition, participants provided an initial prompt and observed 9 candidate images generated by paraphrasing that prompt.
        They could then give textual feedback along with their preferred images. 
        We adopted DSPy~\cite{khattab2023dspy} to refine the user preferred prompts based on their feedback and then paraphrase the prompts for the next iteration.
    \item \emph{Clarification}:
        Participants in this condition began by providing an initial prompt and were then presented with a series of clarification questions designed to make their prompts more specific. 
        They could answer these questions to provide additional details until they were ready to view the generated images. 
        The clarification questions were generated using the method proposed in~\cite{hahn2024proactive}, which has been shown to effectively align image generation results with the user’s intended goals. 
        In each iteration, 9 candidate images were generated based on the previous clarification responses. 
        Participants could select their preferred images, further edit the corresponding prompts, and answer additional clarification questions to refine the results in subsequent iterations.
    \item \emph{\method}:
        Using \method, participants followed the workflow described in~\autoref{sec:workflow}.
        They started with an initial prompt and then selected the generation results aligned with their goal from 9 candidates in each iteration.
        Then our method automatically optimized the prompts and generated new candidates.
        In our study, we hided all prompts from participants, in order to investigate the participants' performance when using \method with minimal effort.
\end{itemize}

To be noted, except for the \emph{PromptCharm} condition, participants were shown 9 image candidates in each iteration, to evaluate how our preference-guided prompt optimization performs compared to existing methods.
For the \emph{DSPy} and \emph{Clarification} conditions, these candidates were generated by paraphrasing the optimized prompt nine times in each iteration, while in \emph{\method}, the candidates came from our three strategies.
The study interface is illustrated as in~\autoref{fig:study_platform}.

In the study, we included two kinds of image generation tasks: close- and open-ended tasks, following prior work~\cite{wang2024promptcharm, feng2023promptmagician}.
In the close-ended tasks, participants were asked to generate an image as similar as possible to a target image, in terms of appearing objects, visual style, composition and so on.
For the open-ended tasks, participants were asked to have their own target in their mind, which could be vague and uncertain, but at least including the essential objects and a few visual details and the general vibe.
Then they were asked to generate an image to achieve their implicit goal.

To control the difficulty across trials and participants, we controlled the number of appearing objects in the generation task.
We first extracted a set including 20 objects that appeared the most from the TIFA160 dataset~\cite{hu2023tifa}.
Then for the close-ended tasks, we randomly sampled 3 objects from the set and asked a LLM to generate a target image prompts by adding descriptors and general visual vibe to these objects.
For the open-ended tasks, we asked participants to include at least 3 objects in their target image.
They could select from the given set or come up with the objects themselves, while keep using different objects across trials.

In each iteration, if participants were satisfied with the current generation results (or with one of outcomes), they clicked "satisfactory" and stopped the current trial, otherwise "see more" to enter the next iteration.
To assess the optimization efficiency, we recorded the minimum number of iterations needed to reach satisfaction in each trial.
We also collected subjective feedback using the NASA-TLX~\cite{HART1988139} and the Creativity Support Index (CSI)~\cite{DBLP:journals/tochi/CherryL14} questionnaire.

\subsection{Procedure}

Participants were first introduced to the study and asked to provide their demographic information.
Then they completed a warm-up trial in each condition to get familiar with the interface.
Once they were ready, they started one trial by providing an initial prompt.
In each iteration, participants were asked to observe the current generation results and decide if they were satisfied with the current results and stop the trial.
If not, participants refined the prompts by: in \emph{PromptCharm}, they edited the prompts using the tools provided; in \emph{\method}, they selected the images that best aligned with their goals; in \emph{DSPy}, they selected preferred images and provided textual feedback to guide prompt refinement; and in \emph{Clarification}, they edited their preferred prompts and answered more clarification questions.
Each trial lasted at most 10 iterations to constraint the overall user study time.

Each participant completed 8 trials (4 conditions $\times$ 2 tasks $\times$ 1 image).
They always began with the close-ended task: after completing all 4 close-ended trials, they proceeded to the open-ended tasks.
We kept this order to present participants with concrete examples first, thereby providing inspiration and reference points for their subsequent open-ended tasks.
The condition order was counterbalanced with a latin-square.
After each trial, participants were asked to fill out the NASA-TLX and the CSI questionnaire.
The study lasted approximately 60 minutes.

\subsection{Apparatus \& Participants}
We recruited a total of 16 participants, with 10 males and 6 females, aged 24 to 28  (\statsum{26.78}{1.21}{}).
The study was conducted on a desktop computer connected to a server running Ubuntu 24.04 with an NVIDIA H200 GPU. 
We used the Stable Diffusion XL model configured with a guidance scale of 7.5 and 50 diffusion iterations.
Each image was generated at a resolution of 1024 $\times$ 1024 pixels, requiring less than 5 seconds for a single image and approximately 20 seconds for a batch of 9 images. 
Participants interacted with the system via mouse and keyboard, with the interface presented in full-screen mode on a 27-inch monitor with a resolution of 2560 $\times$ 1920. 
The underlying language model was Gemini 2.5 Flash.
All prompts are provided in the Supplementary Material.


\subsection{Results}

\begin{figure*}[h]
    \centering
    \includegraphics[width=\linewidth]{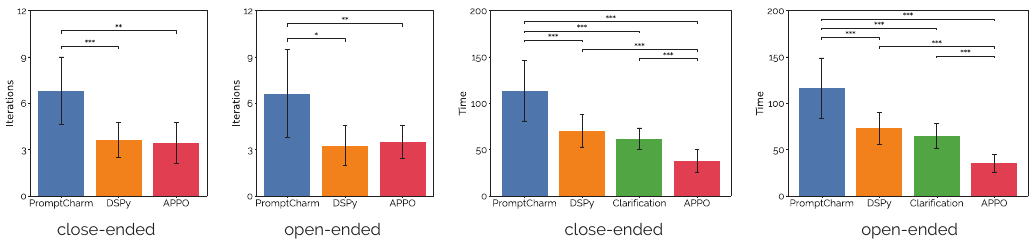}
    \caption{Number of iterations and total time (in seconds) spent required to achieve satisfactory for both close- and open-ended tasks across three conditions (means and standard deviations). (*: $p < 0.05$, **: $p < 0.01$, ***: $p < 0.001$).}
    \label{fig:study_results_iteration}
    \Description{The document presents a comparative analysis of two methods, PromptCharm and DSPy APPO, across two types of tasks: close-ended and open-ended. The comparison is visualized using data for Iterations and Time. In the close-ended category, PromptCharm shows iteration values of 12 , 6 , 9 , and 3 , while DSPy APPO has an iteration value of 12. The bottom visualization compares the Time taken (ranging from 50 to 200) for PromptCharm and DSPy on both close-ended and open-ended tasks, suggesting a performance evaluation in terms of speed and convergence metrics.}
\end{figure*}

We recorded the minimum number of iterations required to obtain a satisfactory outcome, as well as the total time spent.
This is because, in the \emph{Clarification} condition, participants were allowed to ask as many questions as they wanted and could provide substantial information within a single generation iteration. 
Such flexibility makes the iteration counts comparison across conditions unfair and thus, we instead focus on comparing the time used to complete the generation tasks.
The total time excludes the duration of image generation and captures only the time during which participants actively interacted with the systems.
We also collected NASA-TLX and CSI questionnaire responses for all conditions.

We first conducted Shapiro–Wilk tests to assess the normality of the minimum iteration counts and time used. The results indicated no violation of normality (\pvall{>}{.05}). 
Accordingly, we applied one-way repeated-measures ANOVA to examine the effect of \textit{Condition} on the number of satisfactory iterations and total time for both close- and open-ended tasks. 
Post-hoc comparisons were performed using paired t-tests with Bonferroni correction.
For subjective measures, we conducted Friedman tests for each NASA-TLX and CSI metric, followed by Wilcoxon signed-rank tests with Bonferroni correction for pairwise post-hoc analyses.
Full tatistical analysis results are reported in~\autoref{sec:full_statistical_results}.

\paragraph{Participants using \emph{\method} needed fewer iterations and less time to reach satisfactory results compared to all baselines.}
As shown in~\autoref{fig:study_results_iteration}, participants using \emph{\method} reached satisfactory outcomes in fewer than four iterations on average.
In contrast, \emph{PromptCharm} required more than six iterations on average before participants were satisfied.
Although participants using \emph{DSPy} completed a similar number of iterations, they spent substantially more time overall, because they needed to interpret the generated results and craft precise textual feedback.
Participants in the \emph{Clarification} condition similarly spent more time, as they were required to answer multiple clarification questions before seeing each new set of candidates.
Overall, both \emph{DSPy} and \emph{Clarification} required significantly more time than \emph{\method}, even when completing a similar number of iterations.

\begin{figure*}[h]
    \centering
    \begin{subfigure}{0.48\linewidth}
        \centering
        \includegraphics[width=\linewidth]{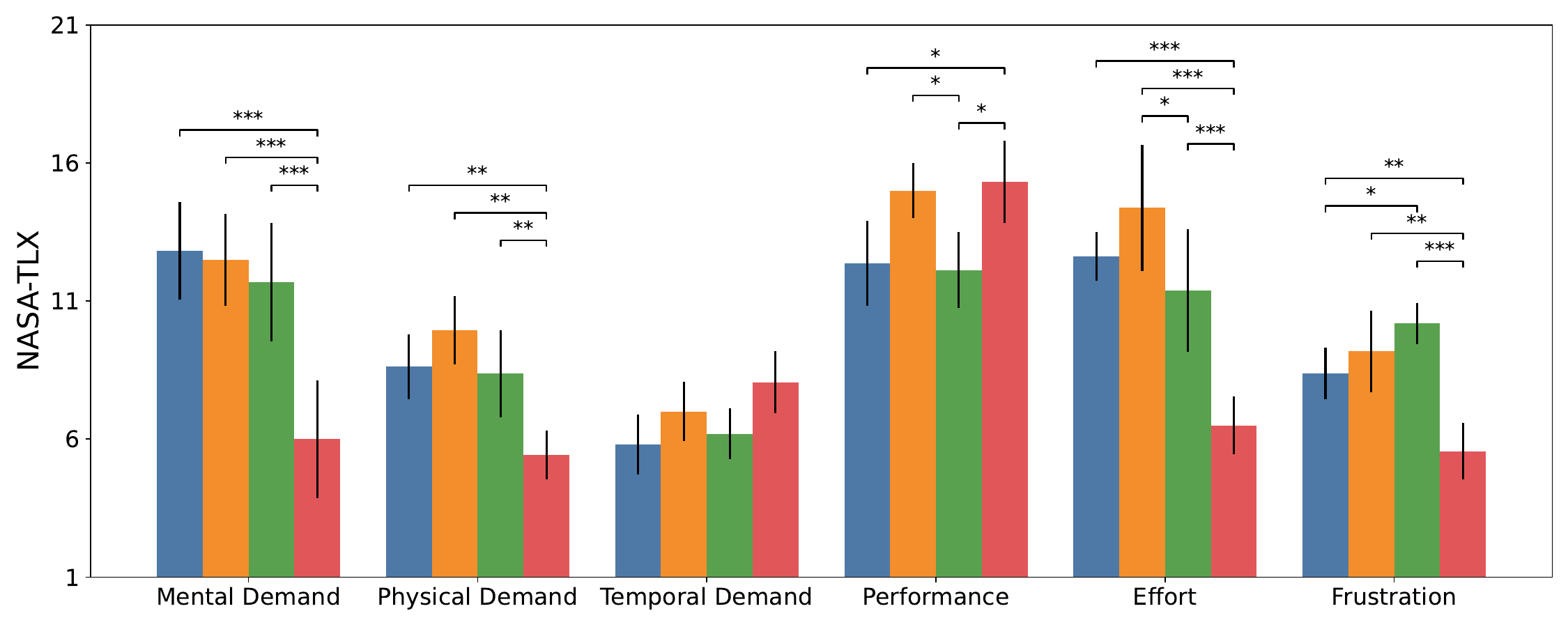}
        \caption{NASA-TLX results for close-ended tasks.}
        \label{fig:study_results_nasa_close}
    \end{subfigure}
    \hfill
    \begin{subfigure}{0.48\linewidth}
        \centering
        \includegraphics[width=\linewidth]{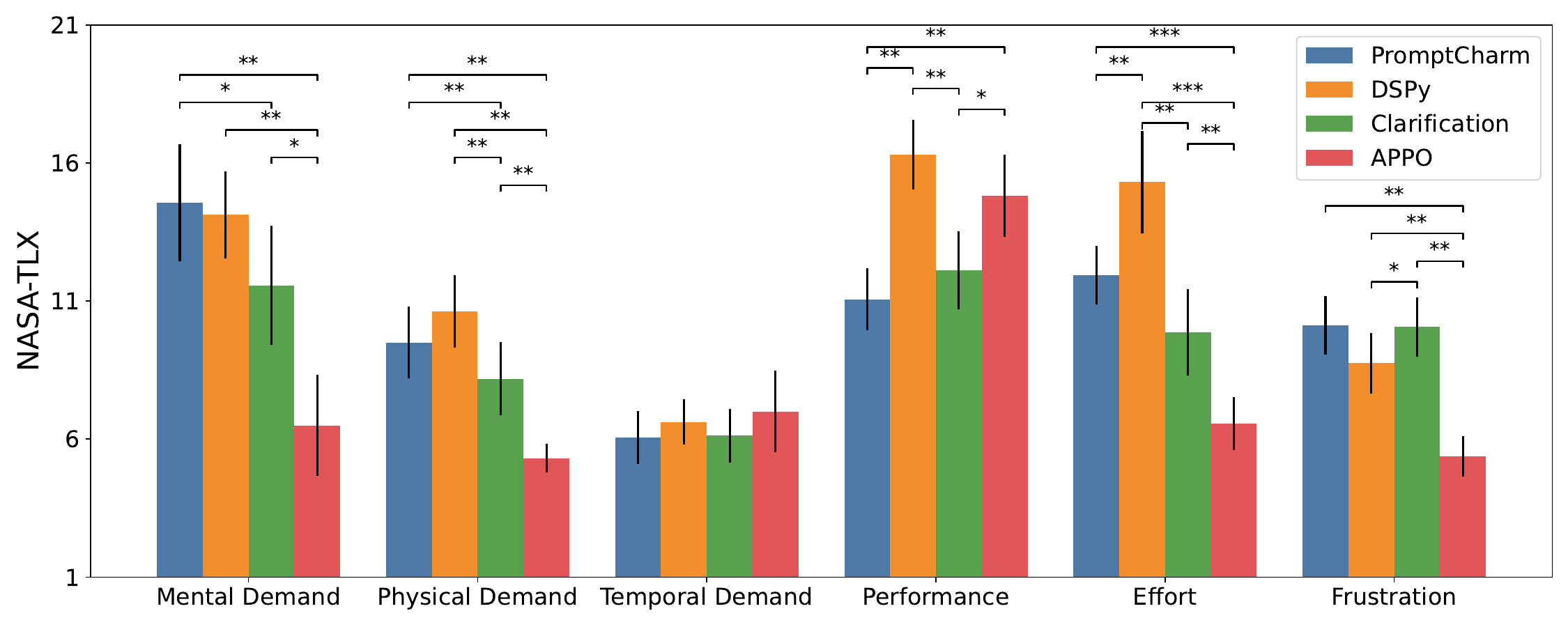}
        \caption{NASA-TLX results for open-ended tasks.}
        \label{fig:study_results_nasa_open}
    \end{subfigure}

    \caption{NASA-TLX questionnaire results for both close- and open-ended tasks across all four conditions (means and standard deviations). (*: $p < 0.05$, **: $p < 0.01$, ***: $p < 0.001$).}
    \label{fig:study_results_nasa}
    \Description{
    (a): The bar chart displays NASA-TLX workload scores, which range from 1 to 21, for six dimensions: Mental Demand, Physical Demand, Temporal Demand, Performance, Effort, and Frustration. Each dimension shows four distinct colored bars, representing different conditions, with error bars indicating variability. Statistical significance is noted above the bars, using asterisks (*, **, ***) to highlight significant differences in scores between the conditions for Mental Demand, Physical Demand, Performance, Effort, and Frustration, while Temporal Demand shows no marked significance between the conditions. The highest average scores across the dimensions are observed in Performance and Effort, while Temporal Demand and Frustration have the lowest overall scores.
    (b): The bar chart illustrates NASA-TLX workload scores, ranging from 1 to 21, comparing four conditions—PromptCharm, DSPy, Clarification, and APPO—across six dimensions: Mental Demand, Physical Demand, Temporal Demand, Performance, Effort, and Frustration. In general, scores for Mental Demand, Physical Demand, Performance, Effort, and Frustration show statistically significant differences between the four conditions, while Temporal Demand scores are closely clustered around 6, showing no marked significance. The conditions DSPy and APPO tend to have the highest scores in Performance and Effort, while APPO consistently shows the lowest scores in Mental Demand, Physical Demand, and Frustration, indicating a relatively low perceived workload in those areas for that condition.
    }
\end{figure*}

\paragraph{Participants experienced lower workload and effort with \emph{\method}, while achieving comparable or better performance compared to all baselines.}
As shown in~\autoref{fig:study_results_nasa}, participants reported significantly lower mental demand, physical demand, and overall effort when using \emph{\method} than when using any of the three baselines. 
They also perceived their task performance to be better than with \emph{PromptCharm} and \emph{Clarification} (for close-ended tasks), and comparable to \emph{DSPy}.
These differences arise because the baseline methods required participants to actively provide textual feedback or refine prompts themselves, which imposed additional mental effort, while \emph{\method} relied on simple interactions with minimal physical and mental demand. 
Participants also felt less frustration with \emph{\method}. 
In contrast, \emph{PromptCharm} required considering multiple refinement strategies that did not always yield better results, and \emph{Clarification} sometimes asked too many clarifying questions, which participants found uncomfortable.
Overall, these findings indicate that \emph{\method} enables participants to achieve comparable or even better performance with significantly less workload and frustration.

\begin{figure*}[!ht]
    \centering
    \hspace{1cm}
    \begin{subfigure}{0.4\linewidth}
        \centering
        \includegraphics[width=\linewidth]{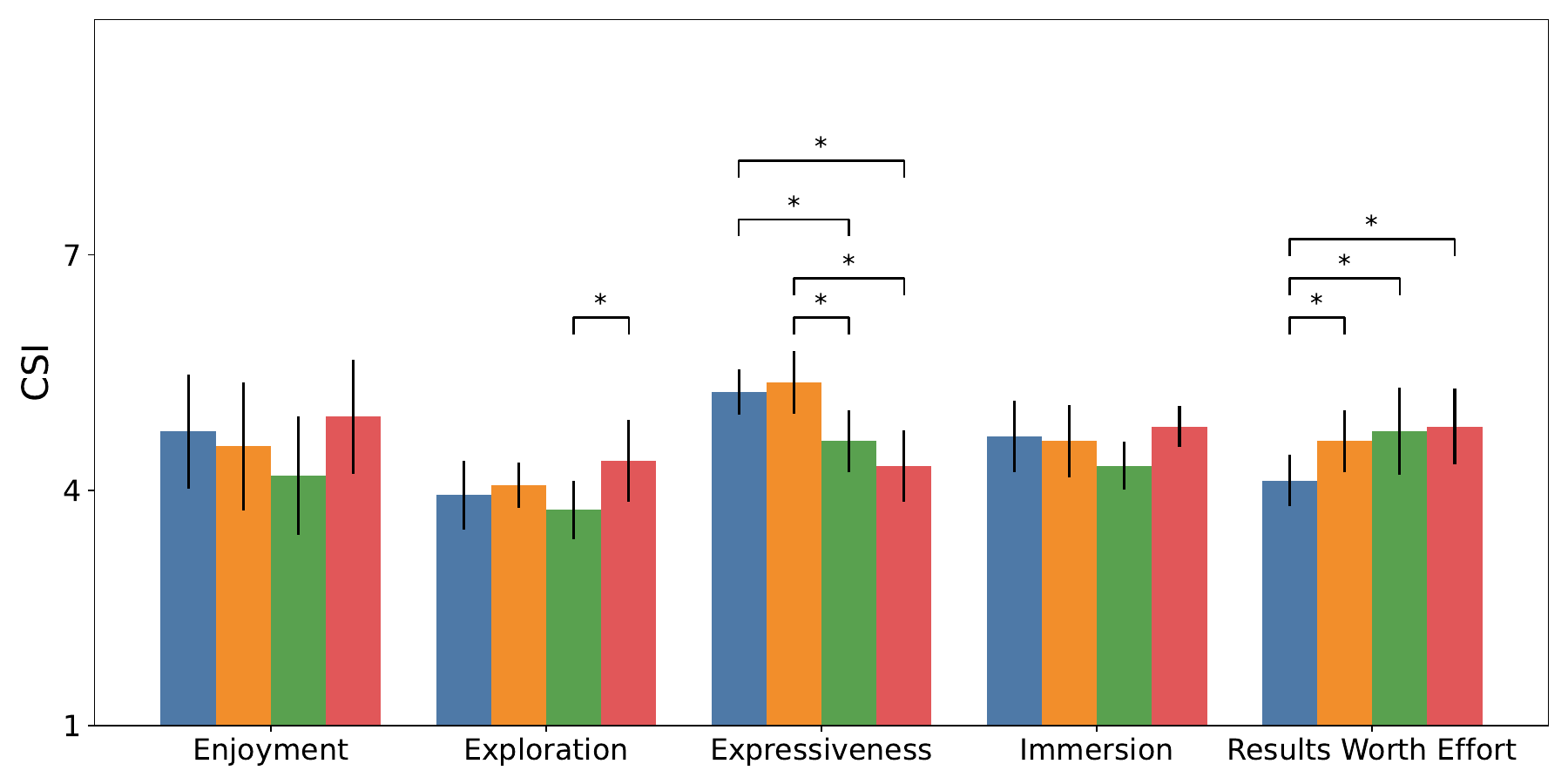}
        \caption{CSI results for close-ended tasks.}
        \label{fig:study_results_csi_close}
    \end{subfigure}
    \hfill
    \begin{subfigure}{0.4\linewidth}
        \centering
        \includegraphics[width=\linewidth]{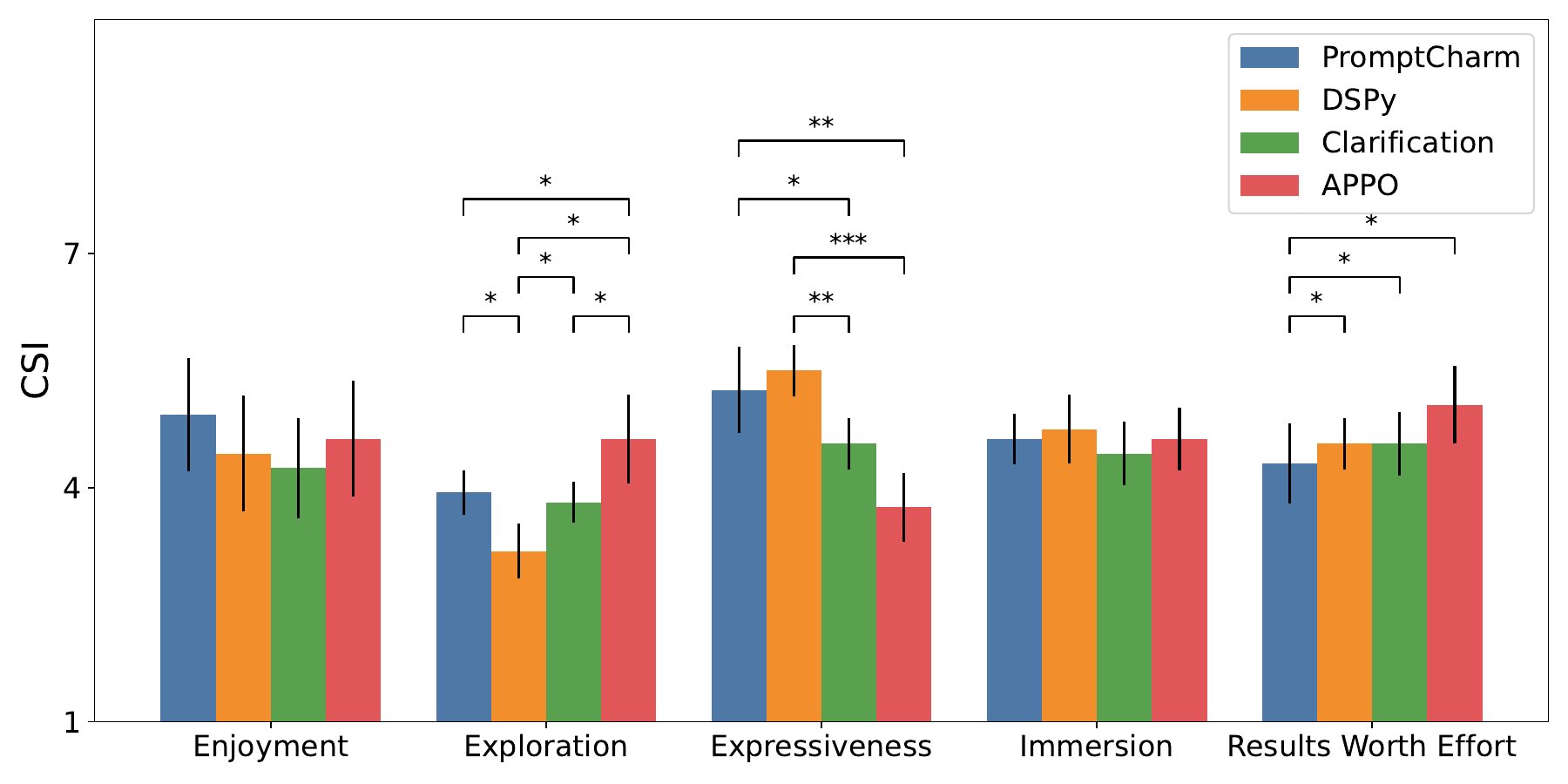}
        \caption{CSI results for open-ended tasks.}
        \label{fig:study_results_csi_open}
    \end{subfigure}
    \hspace{1cm}

    \caption{CSI questionnaire results for both close- and open-ended tasks across all three conditions (means and standard deviations). (*: $p < 0.05$, **: $p < 0.01$, ***: $p < 0.001$).}
    \label{fig:study_results_csi}
    \Description{
    (a): The bar chart displays CSI scores, ranging from 1 to 7, across five dimensions: Enjoyment, Exploration, Expressiveness, Immersion, and Results Worth Effort. Each dimension features four colored bars, representing different conditions, accompanied by error bars. Statistical significance, marked by asterisks (*), is observed for Exploration, Expressiveness, and Results Worth Effort, indicating notable differences between conditions in those dimensions. Expressiveness shows the highest average scores across all conditions, clustering around 5, while Exploration has the lowest average scores, falling below 4.5.
    (b): The bar chart presents CSI scores, ranging from 1 to 7, for four conditions (PromptCharm, DSPy, Clarification, and APPO) across five dimensions: Enjoyment, Exploration, Expressiveness, Immersion, and Results Worth Effort. All five dimensions show some degree of statistically significant differences between the conditions, marked by asterisks, with the most pronounced differences seen in Expressiveness and Exploration. Expressiveness has the highest overall scores, with DSPy achieving the highest average score in this dimension, while Exploration has the lowest overall scores. For Results Worth Effort, the APPO condition has the highest average score.
    }
\end{figure*}

\paragraph{Participants felt their results were more worthwhile with \emph{\method}, with the cost of reduced expressiveness.}
As shown in~\autoref{fig:study_results_csi}, participants rated the worth of their results significantly higher when using \emph{\method} compared to all baseline conditions, suggesting that they perceived \emph{\method} enabling good outcomes with less effort.
However, participants reported significantly lower expressiveness with \emph{\method}. 
This is likely because, in the baseline conditions, they could convey preferences or generation intents through textual feedback or refining the prompts, while \emph{\method} only allowed them selecting preferred images.

\subsection{Discussion}
\label{sec:user_study_discussion}

\paragraph{Preference-only feedback versus richer forms of feedback}
In our study, we compared \emph{\method} with two baselines: \emph{PromptCharm}, a state-of-the-art prompt-engineering tool, and \emph{DSPy}, which refines prompts using users’ textual feedback.
Participants using \emph{\method} achieved comparable or better results with significantly less effort and experiencing lower workload.
These findings reinforce our core motivation: although tools like \emph{PromptCharm} or \emph{DSPy} can help users reach desirable outcomes, they still rely on users to reason about prompt refinements or articulate detailed feedback. 
Consequently, the cognitive load still remains on the human.
In contrast, \emph{\method} requires only binary preference feedback, shifting the reasoning process to the model and achieving similar performance in fewer iterations.

\paragraph{Integrating preferences with richer forms of feedback}
However, \emph{\method} and richer forms of feedback could be compatible. 
It is certainly possible to expose the intermediate prompts to users and allow them to provide richer forms of feedback beyond preferences. 
When issues are obvious or easy to articulate, textual feedback can resolve them more quickly than waiting for the algorithm to infer them. 
Likewise, when the model introduces a new keyword that inspires the user, it may be beneficial to allow users explicitly refining the prompts.
However, this comes with the same challenge faced by traditional prompt refinement workflows: users often lack the knowledge to judge how specific prompt elements influence the generation process. 
Therefore, future research should explore how to combine preference feedback with richer user signals in ways that amplify both strengths while mitigating potential biases or misguidance.

\paragraph{Advantages and disadvantages to present prompts to users}
In our user study, we intentionally did not show optimized prompts to participants, even though our system keeps them readable. 
This is due to we would like to investigate how much \emph{\method} can refine the prompts with minimal user effort. 
In contrast, both \emph{Clarification} and \emph{PromptCharm} required participants to read and evaluate prompt variants in each iteration, which added extra workload to users.
Our study results further revealed that though participants perceived more expressiveness in these conditions, they also reported significantly more workload and took longer to complete the tasks.
These findings show that hiding the prompts does not hinder our method effectively refining the prompts based on participants' preference signal.
While in practical employment, users may still be given the option to inspect prompts when they desired.
For instance, they could learn the vocabulary from the optimized prompts that could  support future generation tasks.

These points further highlight our motivation and contribution in this paper: we aim to reduce user effort while improving outcomes in interactive generation tasks.
By combining efficiency, satisfaction, and creativity support, \method brings generative model closer to everyday creative tasks, where users benefit from quick, inspiring, and satisfying generation without extensive trial-and-error.

\begin{figure*}[h]
    \centering
    \includegraphics[width=\linewidth]{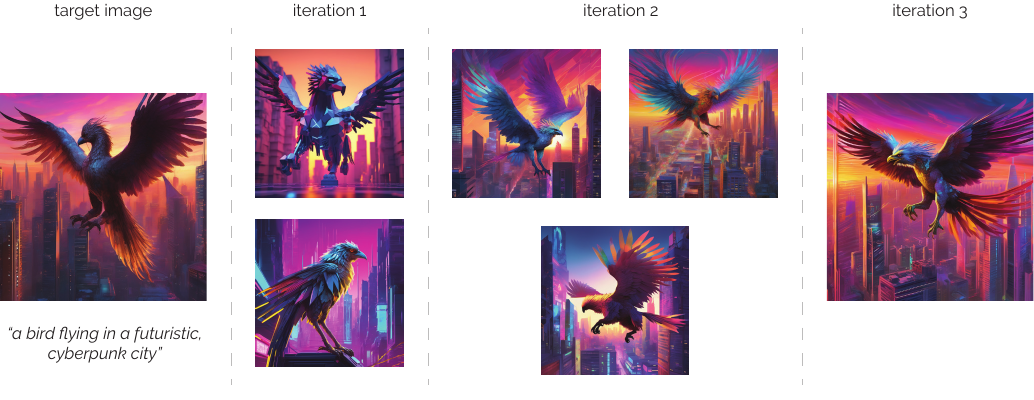}
    \caption{Intermediate results from P3 after the 1st, 2nd, and 3rd iterations using \method on close-ended tasks.
    }
    \label{fig:intermediate_close}
    \Description{The page displays four images generated from the text prompt, “a bird flying in a futuristic, cyberpunk city". The first image is labeled the target image and shows a large, dark bird with its wings fully spread, soaring against a sunset or sunrise over a cityscape of tall, colorful buildings, rendered in a futuristic, vibrant style. The three subsequent images, labeled iteration 1, iteration 2, and iteration 3, are variations of this concept, each depicting a large, mythical-looking bird—possibly a phoenix or eagle—with bright blue and fiery red/orange coloring, flying above a city with high-rise structures, all rendered in the characteristic colorful, neon, and high-tech aesthetic of cyberpunk.}
\end{figure*}

\begin{figure*}[h]
    \centering
    \includegraphics[width=\linewidth]{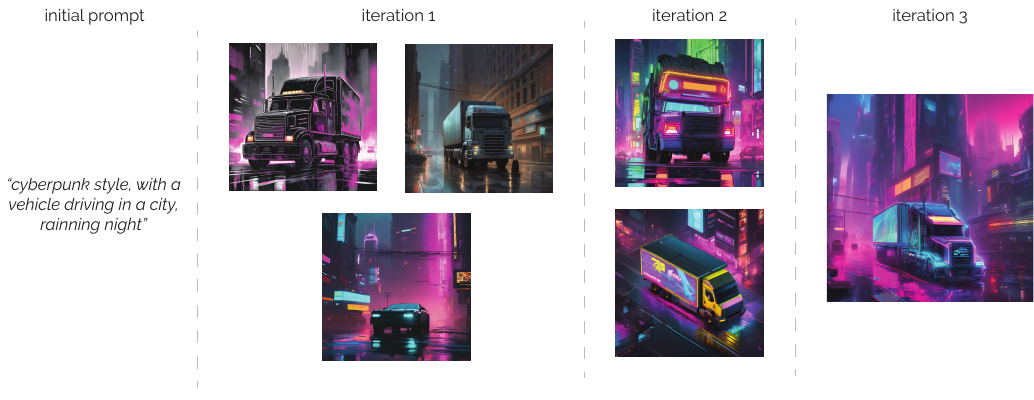}
    \caption{Intermediate results from P10 after the 1st, 2nd, and 3rd iterations using \method on open-ended tasks.
    }
    \label{fig:intermediate_open}
    \Description{The page presents a sequence of four image generation results based on the initial prompt, "cyberpunk style, with a vehicle driving in a city, rainning night". The images are organized into columns for the initial prompt, iteration 1, iteration 2, and iteration 3. The initial prompt column displays a single image of a large, dark semi-truck rendered with neon pink and blue lighting, characteristic of the cyberpunk style. Iteration 1 features two images: one showing a truck similar to the first, and a second showing a smaller car in the same neon-lit, rainy street environment. Iteration 2 also shows two images: a different truck design with glowing yellow accents, and an overhead view of a yellow and black truck on the neon street. The final column, Iteration 3, shows a single, highly stylized image of a semi-truck driving through the intensely vibrant, rainy, and blurred cyberpunk city background.}
\end{figure*}

\paragraph{Intermediate results using \method}
As shown in~\autoref{fig:intermediate_close} and~\autoref{fig:intermediate_open}, we illustrate two example generation processes to demonstrate how \emph{\method} refines prompts using only preference feedback.

In~\autoref{fig:intermediate_close}, the participant’s task was to imitate the target image on the left. 
Although the participant began with a very simple initial prompt, the first iteration already preserved key elements (a bird and cyberpunk city) while exploring diverse visual styles.
For example, the top image depicts a low-poly or robotic bird, whereas the bottom image emphasizes neon lighting in the city.
After the participant selected two preferred images, \emph{\method} refined the prompts. 
By the second iteration, the outputs aligned more closely with the target: the skylight remains consistent across the top two images, the bird becomes more realistic and airborne, and the background city maintains the neon cyberpunk aesthetic. 
The algorithm also explores variations that the user did not specify but are relevant to the target, such as colorful birds and a skyline resembling the target image, despite these details not appearing in the initial prompt.
By the third iteration, the participant selected an image they judged sufficiently similar to the target.

Similarly, in \autoref{fig:intermediate_open}, the participant began with a simple prompt describing a vehicle in a cyberpunk city. 
In the first iteration, the participant selected two images featuring trucks and a third featuring a car in neon lights. 
Notably, all three images shared the characteristic of wet ground, which reflected vehicle shadows and aligning with the “raining night” description.
In the second iteration, \emph{\method} integrated these preferences by generating images of a truck within a cyberpunk setting, illuminated by neon lights. 
However, these images still lacked sufficient city context. 
Through mutation, \emph{\method} expanded its exploration in the third iteration, adjusting the scene’s perspective to reveal more of the urban environment. 
And the participant found this final result satisfactory.
Together, these two examples illustrate how \emph{\method} learns from users' preference feedback and preserves elements that users favor while maintaining diversity across candidates, effectively balancing exploration and exploitation to converge on a desirable outcome.

\paragraph{"Failure" cases using \emph{\method}}
Across the 32 trials completed with \emph{\method}, only two resulted in participants failing to reach a satisfactory image within 10 iterations. 
Both cases occurred in open-ended tasks. 
By collecting participants’ feedback, examining their initial prompts, and their intermediate results, we identified potential causes.
In the first case, the participant did not follow our instruction, which is the key objects must be clearly provided. 
They wanted the generated image to include a specific fantasy animal but could not recall its name. 
They provided only a vague description (P7: "a big creature that has a long, coiled body, sharp teeth, and scary eyes" but he actually wants "Jörmungandr"). 
While other aspects of the image matched the expectation and \emph{\method} explored several types fantasy creatures across iterations, the exact animal the participant's in mind was never correctly generated. 
This is reasonable as the key object was never correctly provided in the first place, violating our assumption of the system.
In the second case, the participant (P4) included many distinct features that would like to have in the generation outcomes.
And he/she encountered a promising image early on but also selected several alternatives that contained additional features they hoped could be combined later. 
However, these selected candidates also included features that were not part of the participant’s desired set and conflicted with the wanted features. 
This made it difficult for \method to determine which parts of the prompts influenced the participant’s selections. 
As a result, it failed to find the correct combination of features within the limited interaction budget.
These examples highlight potential limitations of \method.
First, it could struggle when users have highly specific intents that they cannot describe precisely.
Second, including too many features makes it hard for \method to identify the correct combination within the limited interaction budget.
In such cases, allowing the users to review the current prompts and further edit can effectively address the challenge. 
APPO can easily be extended to offer this feature, as our prompt search remains comprehensible and directly editable by human users.

\begin{figure*}[h]
    \centering
    \includegraphics[width=\linewidth]{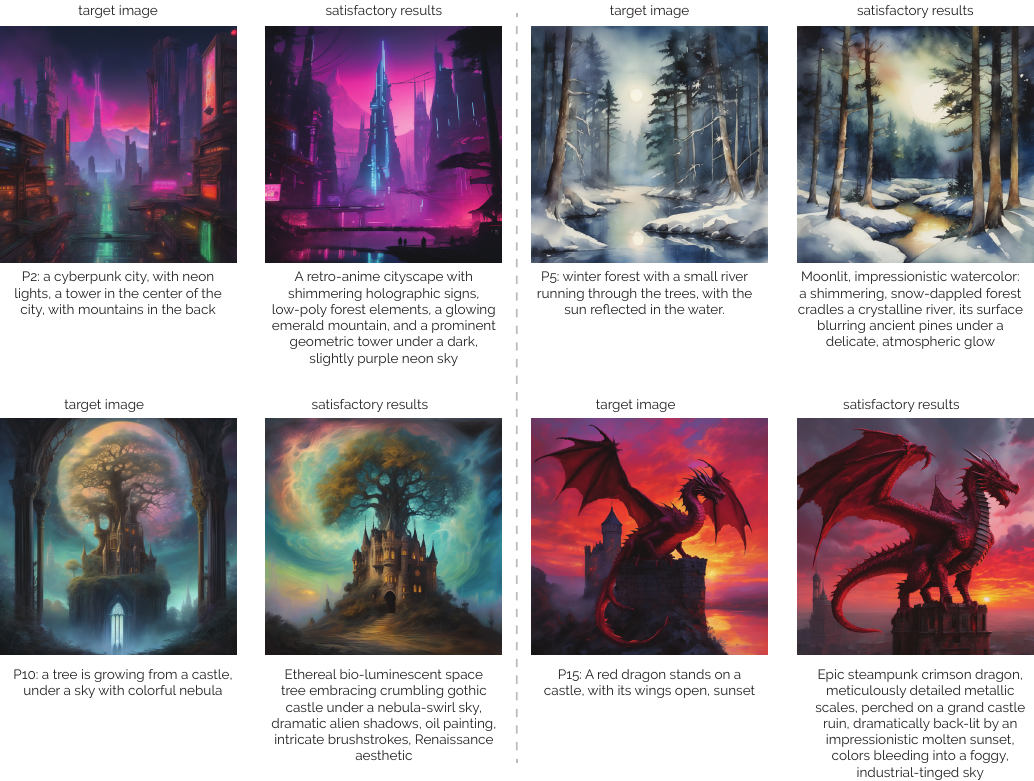}
    \caption{Example of generation results in close-ended tasks by participants using \method.
    }
    \label{fig:close_example}
    \Description{This page displays four pairs of images, each pair consisting of a target image and a second image labeled satisfactory results. The four concepts shown are: a cyberpunk city with neon lights and a central tower , a winter forest with a small river , a large tree growing from a castle under a nebula-like sky , and a red dragon standing on a castle at sunset. For the cyberpunk city, the target image shows a futuristic purple and pink cityscape with a glowing green river, while the second image is a close-up, highly detailed version of a central spire with blue glowing lines. The winter forest images are rendered in a watercolor style, depicting a snowy forest and a reflective stream. The castle/tree images show a large tree above a castle, once framed by pillars and the other with a dark, colorful sky. Finally, the dragon images feature a red dragon perched on a fortress against a bright red and orange sunset.
}
\end{figure*}

\begin{figure*}[h]
    \centering
    \includegraphics[width=\linewidth]{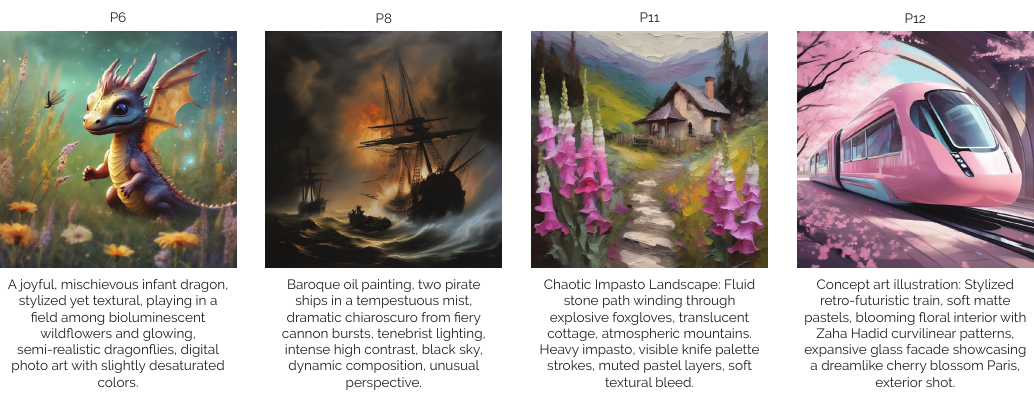}
    \caption{Example of generation results in open-ended tasks by participants using \method.
    }
    \label{fig:open_example}
    \Description{The page displays four distinct images, each generated from a descriptive prompt. The first image (P6) is a stylized yet textural digital photo art piece showing a joyful, mischievous infant dragon playing in a field among glowing, bioluminescent wildflowers, with a semi-realistic dragonfly in the air. The second image (P8) is a dark, dramatic Baroque oil painting that uses intense high contrast and tenebrist lighting to depict two pirate ships caught in a tempestuous mist, illuminated by a fiery cannon burst. The third image (P11) is a Chaotic Impasto Landscape, rendered with heavy knife palette strokes, featuring a stone path winding past explosive pink foxgloves toward a translucent cottage set against atmospheric mountains. The final image (P12) is a concept art illustration of a stylized, retro-futuristic pink train in soft matte pastels, showcasing an exterior shot of the train passing through a dreamlike cherry blossom setting.}
\end{figure*}

We also present example generation results from the user study using \method in~\autoref{fig:close_example} and~\autoref{fig:open_example}.

%% file: sections/7_discussion.tex
\section{Discussion}

\method enables users to guide the generation process with their preferences by automatically optimizing prompts based on prior feedback.
Essentially, \method adopts three strategies: retainment, alignment, and expansion, which together enable the optimization explores users' potentially satisfactory regions in the broad prompt space by learning from their sparse but high-quality feedback.
To escape local optima suggested by preference signals, \method employs an adaptive mechanism that balances the three strategies. 
This ensures both effective exploitation and active exploration across iterations, ultimately achieving better results in fewer steps.
In doing so, \method not only addresses the challenge that users often lack the right vocabulary to refine prompts, but also adapts to their evolving goals by continuously learning from feedback while maintaining active exploration.

\paragraph{Extending Preference-Guided Optimization to Other Generation Tasks}
While this paper focuses on image generation, the core principles of our preference-guided prompt optimization are potentially applicable to other generative domains, including text generation~\cite{radford2019language, brown2020language}, code synthesis~\cite{fried2022incoder, jiang2024survey}, and music generation~\cite{huang2018music}. 
Conceptually, the interactions will remain the same for other applications: generate candidate outputs from an initial prompt, collect preference feedback, and iteratively refine the prompt using our strategies and adaptive policy. 
Extending to these domains may require task-specific adjustments, such as designing appropriate candidate representations or feedback interfaces.
Moreover, some generation outcomes (such as videos) may involve richer or more complex information which can be difficult to fully articulate in a single prompt. 
This introduces additional challenges that are not present in image generation tasks, as discussed below.

\paragraph{Extending Preference-Guided Optimization to Complex Generation Tasks}
In this paper, we evaluated our method on tasks where users could provide an initial prompt containing the key information necessary for generation. 
Our results show that preference-guided optimization can effectively identify relevant descriptors and modifiers, but recovering missing key elements (e.g., the primary object in an image) remains challenging due to the vast search space and the sparsity of preference signals.
This limitation raises concerns for more complex tasks. 
First, the assumption that users can specify all essential elements upfront may not hold. 
For example, designing a webpage with numerous components and intricate layouts is difficult to capture in a single prompt, and designers typically refine iteratively while observing intermediate results. 
Second, the assumption that the optimizer can reliably infer the correct modifiers or descriptors becomes less feasible as the space of possible descriptors grows, such as when generating entire 3D environments—potentially requiring many more iterations or making optimization impractical.
Furthermore, when users provide additional information mid-process, prior optimization experience may no longer apply, forcing the system to restart and increasing frustration and cognitive load. 
More broadly, extending preference-guided optimization to support tasks that require iterative user input and evolving goals remains an open challenge and a critical direction for future work.

\paragraph{Toward multi-modal prompt input}
Textual prompts remain the most common and primary way for users to interact with generative models, so we focus on optimizing textual prompts to support a wide range of tasks that rely solely on text. 
However, modern generative models increasingly support multi-modal inputs. 
For instance, recent models such as DALL·E 3 accept both text and images to better steer the generation process. 
Our workflow can naturally extend to these scenarios: the fundamental principle of APPO lies in contrasting preferred and non-preferred outcomes and generating new variations.
Modern models (e.g., VLMs) have similar ability for such analysis across modalities. 
Nevertheless, multi-modal prompts introduce new challenges, such as determining how to balance and combine signals of modalities, how to preserve essential elements when modalities carry overlapping or conflicting information, and how to design interfaces that let users express preferences efficiently across different types of media. 


\paragraph{Leveraging Shared Preferences for Efficient Optimization}
While preferences in generative tasks are highly individual and context-dependent, they often exhibit recurring patterns across users and tasks. 
We can potentially leverage the accumulated optimized prompts and their corresponding outputs to improve the efficiency and quality of future generations. 
This idea aligns with the concept of meta-learning, or ``learning to learn'' in machine learning~\cite{li2025efficient, Liao2026HOMI, liao2024meta, 10.1145/3706598.3713603}.
Although optimized prompts inevitably diverge across tasks and users, reusing them as a population source for initialization can accelerate optimization. 
For example, when a user specifies a keyword or descriptor (e.g., comic-like), the system could query a repository of previously optimized prompts containing similar descriptors and integrate them into the expansion strategy. 
Similarly, user-specific tendencies can be captured: a designer who consistently favors a particular color theme will have optimized prompts that encode effective text–modifier combinations for obtaining that theme. 
By reusing and recombining such prompts, the optimizer can reduce redundant exploration and converge more quickly to satisfactory outcomes.
Future work should investigate maintaining a bank of optimized prompts together with a query mechanism to identify relevant prior prompts for reuse. 
Such mechanisms would allow optimization to start from a stronger baseline, improving both efficiency and personalization, while also raising new questions about how to balance prior knowledge with the flexibility to adapt to novel user goals.


\paragraph{Preference Shifts and Re-Optimization.}
Users' preferences can evolve even within a single generation task, particularly as they explore different outputs as they might have a more clear goal when seeing various outcomes.
These shifts imply that earlier feedback may no longer reflect the user's current intent, which can reduce optimization effectiveness if all past feedback is treated equally.
To address this, future prompt optimization methods could incorporate strategies that dynamically adjust the influence of prior observations. 
For example, the prompt optimizer could prioritize more recent feedback or apply techniques such as sliding windows, exponential decay, or other adaptive forgetting mechanisms. 
These strategies allow the system to remain aligned with the user’s evolving preferences while still benefiting from earlier iterations.
Integrating such mechanisms into our prompt optimization framework would make it more responsive and robust, enabling generative models to adapt in real time as users refine their goals or explore new directions. 
This could make our method balances efficiency with flexibility, ensuring that optimization remains effective even in dynamic or exploratory tasks.

\paragraph{Other Approach for Preference-Guided Prompt Optimization}

In this paper, we propose \method, which iteratively refines prompts based on users’ preference feedback to align them with their generation intentions. Prior work has explored incorporating human feedback into prompt optimization in a different direction:  training a separate model for prompt selection. 
Specifically, APOHF~\cite{lin2024prompt} trained a neural network on simulated pairwise comparisons to predict the goodness of prompts and then selected candidates from a predefined database. 
A fundamental constrain is the fixed prompt database: it cannot synthesize new prompts beyond the database, making it difficult to adapt to novel or highly personalized goals. 
This limitation reduces its applicability in real-world scenarios, where user intentions are inherently open-ended and often not represented in a predefined set.
Future work should overcome this limitation by developing a generative prompt optimizer that learns directly from preference signals. 
Rather than being restricted to existing prompts, such a model could generate open-ended, preference-aligned candidates. 
This would combine the scalability of large-scale preference learning with the adaptability needed to handle diverse and evolving user goals in practice.


%% file: sections/8_conclusion.tex
\section{Conclusion}

We present \method, a human-centered prompt optimization method that allows users to guide image generation process using simple selection-based preference feedback. 
\method iteratively optimizes the prompts based on the the preference feedback, which lead to improvement in the generation outputs, gradually converging toward the user's implicit goals.
To address the fundamental trade-off between exploration and exploitation in the vast prompt space, we introduce three complementary strategies: retainment, alignment, and expansion. 
These strategies, together with an adaptive exploration policy, support \method avoid local optima and converge efficiently toward high-quality prompts.
We evaluated \method through ablation and user studies on both target-matching and open-ended image generation tasks. 
Results show that it reduces the number of iterations and cognitive effort compared to manual editing or existing optimization methods, while maintaining high user satisfaction and control over the outputs.
These findings demonstrate that our preference-guided optimization method can effectively refine prompts via the simplest form of interaction, binary selections, bridging the gap between human intent and automatic prompt adjustment. 
We envision this work as a step toward a new form of human–AI collaboration, where AI not only translates human ideas into outputs but, from the smallest yet meaningful preferential signals, actively aligns with human goals and discovers the language to guide the generation.



%% file: sections/5_synthetic.tex
\section{Synthetic Tests}

We present \method, a framework that enables effective prompt optimization for image generation tasks purely from binary human preferences. 
While our user study shows promising results, several open-ended questions remain that cannot be fully addressed without a controlled experiment. 
In practical image generation scenarios, there are different strategies for providing initial prompts and the users may have different levels of target granularity in mind.
Yet, these cannot be systematically examined in a user study, since human prompting strategies and their implicit goals are inherently noisy and difficult to control. 
Moreover, \method integrates multiple strategies and an adaptive policy, but the level of support of each cannot be tested individually within the constraints of a user study. 
To more thoroughly evaluate \method’s ability to infer users’ generation intents and optimize prompts, we further compare it against existing automatic prompt optimization methods that do not incorporate human feedback.
To do so, we conduct synthetic tests on image generation using simulated users who provide well-controlled initial prompts and reliable preference feedback.
Specifically, these are our goals for the synthetic tests:


\begin{enumerate}
    \item \textbf{Validate the effectiveness of \method under different scenarios:} We assess whether \method can offer consistent performance with different \textit{initial prompts} and for different \textit{granularity levels of the targets}.
    \item \textbf{Compare performance without individual strategies:} We evaluate variants of \method with alignment or expansion removed to understand the resulting performance differences.
    \item \textbf{Evaluate the adaptive expansion policy:} We compare the adaptive expansion policy against a non-adaptive baseline to show its impact on convergence speed and final output quality.
    \item \textbf{Compare \method with existing automatic prompt optimization methods:} 
    We compare \method against two state-of-the-art approaches that aim to align generation results with user intent or preferences. 
    One method relies on self-reasoning over previous observations to refine prompts~\cite{yuksekgonul2024textgrad}, while the other generates checklist questions based on prior preference feedback to guide subsequent prompt refinement~\cite{cook2024ticking, viswanathan2025checklists}.
\end{enumerate}

\subsection{Design}
\label{sec:synthetic_design}
We first sample prompts from established, widely used datasets~\cite{hu2023tifa} to obtain a set of unique base prompts. 
Each base prompt is then expanded with additional details to construct a synthetic target prompt; this step also generates different granularity levels of target prompts. 
The target prompt represents the ultimate, implicit goal of the synthetic user, but it is \textbf{not} directly provided as the initial prompt. 
This setup simulates the ambiguous and hard-to-articulate nature of realistic textual instructions. 
Instead, we extract the key information (i.e., the obvious, explicit, and easy-to-formulate elements) from the target prompt to form the initial prompt; this step further leads to different initial prompting strategies. 
This mimics how humans can clearly articulate some parts of their intent but not the full specification.

These initial prompts are then given to different variants of \method (we will refer to them as variants in the following content) and two state-of-the-art prompt optimization methods to start iterative refinement driven by preferences. 
At each iteration, the synthetic user selects $N$ outputs that are most similar to the target prompt, simulating preferential feedback; that is, users choosing options that align with their implicit goals. 
To determine the similarity between each generation output and the target prompt, we employ the CLIP model~\cite{DBLP:conf/icml/RadfordKHRGASAM21} to encode outputs (not the candidate prompts) and target prompt into vector embeddings. 
CLIP is a suitable model particularly because of its capability to map both texts (e.g., target prompt) and images (e.g., generated images) into the same embedding space, enabling computing the alignment across modalities. 
We then use the cosine value between their embeddings to indicate their similarity.
Since different variants lead to different candidates, the iterative process diverges across methods. 
We track the similarity between the selected outputs and the target prompt at each iteration, where higher similarity indicates better performance. 

Below, we expand details of the overall synthetic experiment design.

\paragraph{Scenarios (initial prompt strategies and granularity levels of the target):}

As mentioned, we aim to simulate different scenarios of preference-guided generative processes. 
Two key factors determine these scenarios: 
(1) The strategy for creating the initial prompt: different users may provide varying levels of detail at the starting point.
(2) The granularity of the target: different users or contexts may require goals with differing levels of specificity.

For every task, a pair of initial and target prompts are derived from the same base prompt sampled from existing datasets. 
The base prompt is varied into two types of initial prompts:
A \textbf{complete initial prompt}, which uses the base prompt directly, and a \textbf{keyword-based initial prompt}, which is a simplified version obtained by removing descriptors and extracting key terms using Rapid Automatic Keyword Extraction (RAKE)~\cite{rose2010automatic}.
In parallel, the base prompt is transformed into two types of target prompts:
A \textbf{low-granularity target}, which specifies only an overall style or ``vibe'' without detailed attributes.
A \textbf{high-granularity target}, which specifies modifiers for individual objects and fine-grained visual features.
We performed the transformations with a language model, using the prompt provided in~\autoref{sec:ablation_prompt}.

We systematically combine these variations to create four scenarios:
\begin{enumerate}
\item Complete initial prompt + low-granularity target prompt,
\item Complete initial prompt + high-granularity target prompt,
\item Keyword-based input prompt + low-granularity target prompt,
\item Keyword-based input prompt + high-granularity target prompt.
\end{enumerate}


\paragraph{Synthetic users:}

An important role in the synthetic tests is the synthetic users, whose actions include (1) providing the initial prompts, (2) iteratively comparing the similarity levels between each generated output and the target prompt, and (3) selecting top $N$ similar outputs. 
We have provided how the initial prompts are generated and how the similarity levels are computed. 
In every iteration, the value of $N\in [1,4]$ is randomly sampled from uniform distribution. 


\paragraph{Shared apparatus}

All experiments were conducted on Ubuntu 24.04 using a single NVIDIA H200 GPU. 
We used Gemini 2.5 Flash to generate candidate prompts and Stable Diffusion XL~\footnote{\url{https://huggingface.co/stabilityai/stable-diffusion-xl-base-1.0}} model, configured with a guidance scale of 7.5 and 50 diffusion iterations, to generate images.  
We define a task as a selection of a pair of initial prompt and a target prompt; each task runs for 15 optimization iterations, allowing us to observe how variants behave over iterations.


%

\subsection{Variants of \method and Baselines}

To investigate the contribution of individual strategies and components in \method, we compare it against a set of controlled variants and two state-of-the-art automatic prompt optimization methods. 
As introduced earlier, \method is driven by three strategies (\emph{retainment}, \emph{alignment}, and \emph{expansion}), and it relies on an \emph{adaptive policy} to control exploration intensity. 
By selectively removing or simplifying these components, we can individually investigate their respective contribution.

\paragraph{\method}
The complete \method integrates retainment, alignment, and expansion, with adaptive control over exploration intensity. This enables the optimizer to dynamically adjust its search strategy based on progress.

\paragraph{No-Alignment}
This variant removes the \emph{alignment} strategy while retaining \emph{retainment} and \emph{expansion}. 
Preferred prompts are mutated to generate new candidates, simulating a purely evolutionary process without explicit guidance.

\paragraph{No-Expansion}
This variant removes the \emph{expansion} strategy while retaining \emph{retainment} and \emph{alignment}. 
It summarizes the textual gradient to refine prompts but lacks the ability to introduce random prompt mutations.

\paragraph{No-Adaptation}
This variant removes the \emph{adaptive expansion policy}, eliminating deliberate control over exploration and exploitation behaviors.

\paragraph{Paraphrase}
A minimal baseline in which the prompt is paraphrased into a different full sentence at each iteration, without optimization or incorporating preference feedback. 
This reflects the strategy of producing only complete sentences without guided optimizations.

We further include two baselines to better demonstrate the effectiveness of \method.

\paragraph{TextGrad}
TextGrad~\cite{yuksekgonul2024textgrad} has been shown to effectively align prompts toward outputs that are preferred or receive higher scores.
It estimates a textual gradient from the evaluation results and then use the gradient to refine the prompt.
In our setting, it provides a natural baseline: the method first diversifies the initial prompt into multiple candidate prompts, then infers the user’s generation intent by comparing preferred and non-preferred outputs.
Based on these comparisons, it used the relative preferences to refine the prompts in subsequent iterations.

\paragraph{Self-TICK}
Self-TICK~\cite{cook2024ticking} is a state-of-the-art method for aligning LLM outputs with user instructions by generating checklist questions and applying them as constraints for generation.
For our prompt optimization setting, we adapt Self-TICK by deriving checklist questions from the comparison between preferred and non-preferred results. 
These checklist items are then used to refine the prompt in subsequent iterations, guiding it toward the intent revealed by previous observations.

\subsection{Comparing Variants of \method}


We randomly sample 20 prompts from the TIFA160 dataset~\cite{hu2023tifa} to be the base prompts. 
This dataset provides diverse, natural prompts that reflect realistic user intent, making it well-suited for iterative prompt refinement.  
As detailed in \autoref{sec:synthetic_design}, we include four scenarios in this task, leading to two types of initial prompts and two granularity levels of targets. 
In the \emph{Expansion} strategy of the method variants, we generate five child prompts from each pair of population prompts, and each child prompt is mutated twice.

\subsubsection{Results}  

\begin{figure*}[!ht]
    \centering
    \includegraphics[width=\linewidth]{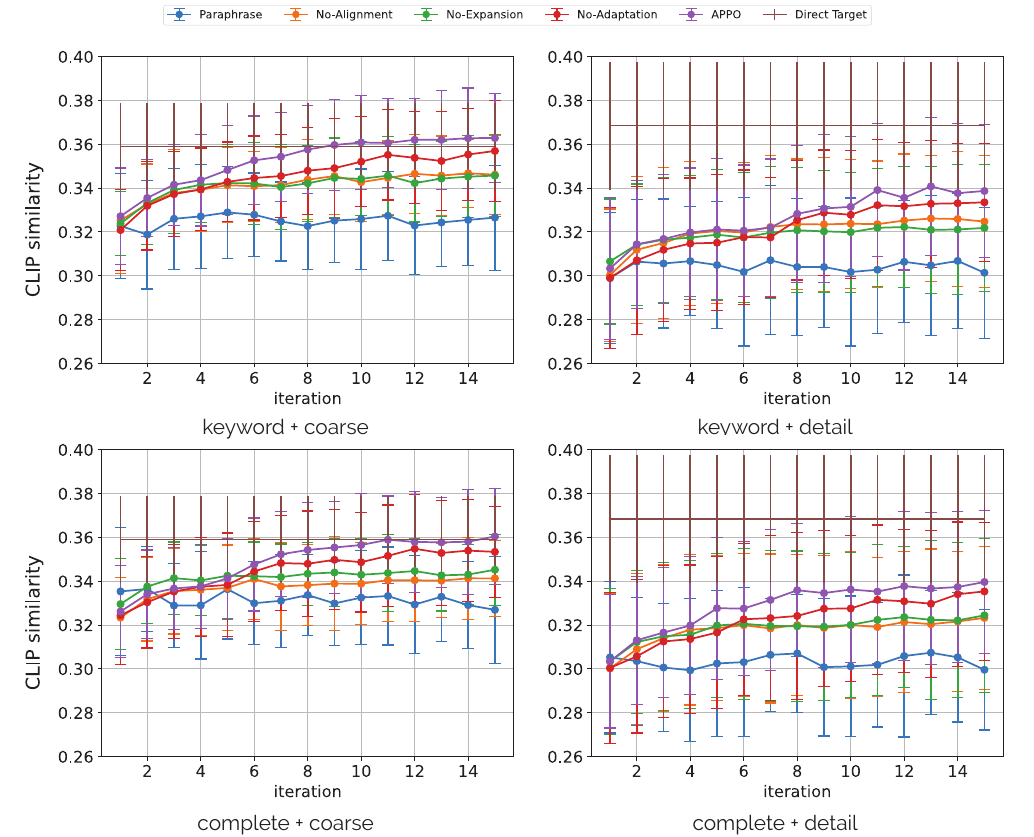}
    \caption{
        Ablation study results using variants of \method.
        We also directly use the target prompt in each scenario to generate results and compute the similarity between the target prompt and its own generated output, serving as the upper bound for the optimization, noted as ``Direct Target.''
        As shown in the figure, when the target prompt is low-granularity, \method refines the prompt to achieve results comparable to this upper bound.
        When the target prompt is high-granularity, however, \method yields slightly lower similarity scores because the target prompt includes multiple descriptors that are missing in the initial prompt and they are difficult to fully recover through the optimization process.
        These results demonstrate that \method effectively discovers prompts that produce images aligned with the target under diverse input–target settings. 
    }
    \label{fig:ablation_image}
    \Description{The image contains four comparative line graphs in an ablation study, all charting CLIP Similarity scores (ranging from 0.26 to 0.40 on the Y-axis) as they evolve across optimization Iterations (ranging from 2 to 14 on the X-axis). The top-left graph compares the performance of the full APPO method against its ablative components, including No-Alignment, No-Expansion, No-Adaptation, Paraphrase, and a Direct Target baseline, demonstrating that the full APPO consistently achieves the highest CLIP similarity over iterations. The remaining three graphs further explore prompt engineering by comparing performance based on prompt structure, specifically contrasting results for keyword + coarse versus keyword + detail, and complete + coarse versus complete + detail, to show how different prompt strategies impact the final image-text alignment score.
}
\end{figure*}

\autoref{fig:ablation_image} shows the simulation results using variants of \method.

\paragraph{Goal 1: Validating the effectiveness of \method under different scenarios}
Across all scenarios, \method demonstrates clear improvements over iterations. 
While \emph{No-Alignment} and \emph{No-Expansion} also improves the results throughout iterations, \emph{\method} still consistently outperforms the other variants.
This demonstrates the overall effectiveness of \method in improving generation performance.
This highlights its generalizability across different initial prompting strategies and target granularity levels.


\paragraph{Goal 2: Comparing the performance without individual strategies}

We compare \emph{No-Adaptation} with \emph{No-Alignment} and \emph{No-Expansion} to evaluate the role of each strategy. 
Both \emph{No-Alignment} and \emph{No-Ex-} \emph{-pansion} show initial gains in similarity during the first 5 iterations but quickly plateau, indicating stuck in local optima. 
This demonstrates that relying on a single strategy is insufficient for efficiently discovering high-quality prompts. 
In contrast, \emph{No-Adaptation} continues to improve until around iteration 10–12, suggesting that the combination of alignment and expansion supports more effective prompt optimization. 
Further analysis shows that \emph{No-Alignment} exhibits consistently higher variability, with larger standard deviations, reflecting active exploration—but it remains unstable and fails to reach a global optimum.

\paragraph{Goal 3: Evaluating the adaptive expansion policy.}
To evaluate the adaptation policy, we compare \emph{\method} with \emph{No-Adaptation}. 
While both perform similarly in the early iterations, \emph{\method} begins to outperform \emph{No-Adaptation} after about 6–8 iterations. 
This advantage arises because the similarity estimates become more stable after several iterations, enabling the adaptive policy to adjust expansion intensity more effectively. 
As a result, \emph{\method} converges to higher-quality final solutions, demonstrating the value of the adaptive expansion policy.

\subsection{Comparing \method to Baselines}

We used the same prompts as in the previous tests and included four scenarios, resulting in two types of initial prompts and two levels of target granularity.
In the \emph{Self-TICK} condition, we generated five checklist questions in each iteration and refined the prompt for up to three refinement rounds or until the prompt satisfied all checklist items.

\subsubsection{Results}

\begin{figure*}[!ht]
    \centering
    \includegraphics[width=\linewidth]{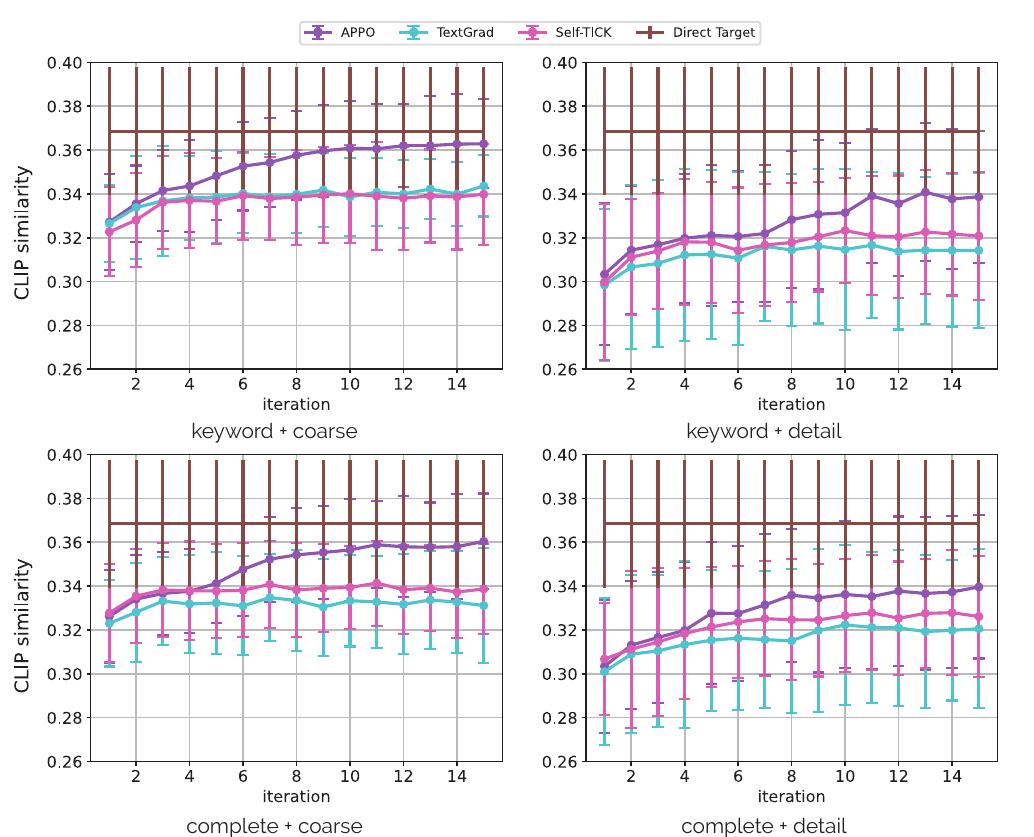}
    \caption{
        Ablation study results comparing \method with TextGrad and Self-TICK.
        We also include a “Direct Target” condition to indicate the upper bound of achievable performance.
        As shown in the figure, across all four scenarios, \method consistently outperforms both TextGrad and Self-TICK.
        \method achieves better results in the early iterations and converges to better final outcomes.
        These results demonstrate that \method outperforms existing automatic prompt optimization methods purely based on preference feedback for image generation tasks.
    }
    \label{fig:ablation_results_baselines}
    \Description{The image contains two line graphs comparing the performance of four different prompt optimization methods—APPO, TextGrad, Self-TICK, and a Direct Target baseline—in terms of CLIP Similarity (ranging from 0.26 to 0.40 on the Y-axis) across optimization Iterations (ranging from 2 to 14 on the X-axis). The top-left graph shows the methods' performance for keyword + coarse prompts, where APPO consistently maintains the highest CLIP score, closely followed by TextGrad. The bottom-left graph shows the results for complete + coarse prompts, where APPO and TextGrad again perform the best, though the separation between the methods is more pronounced than in the keyword prompts.}
\end{figure*}

\autoref{fig:ablation_results_baselines} shows the simulation results using \method and other two prompt optimization methods.

\paragraph{Goal 4: Comparing \method with existing automatic prompt optimization methods}
We compare \method against \emph{TextGrad} and \emph{Self-TICK}.
While all methods perform similarly in the first few iterations (approximately 3–4 iterations), \method continues to make improvements afterward, while both baselines quickly converge and fail to further refine the results.
As a result, \method ultimately achieves better final outcomes than both baselines.
These results demonstrate that \method achieves better exploration and exploitation with various strategies and adaptive policy, leading to more effective outcomes.

\subsection{Summary}

The synthetic tests complement our user study by isolating factors that are difficult to control with real users and by enabling a direct comparison between \method and state-of-the-art automatic prompt optimization methods.
Together, the results address our research goals.
First, \method consistently improves output quality across different initial prompt strategies and target granularity levels, demonstrating robustness and generalizability. 
Second, the ablation studies reveal that the combination of alignment and expansion is essential for efficiently identifying high-quality prompts. 
Third, the adaptive expansion policy proves critical for later iterations, enabling \method to converge to superior final solutions compared to non-adaptive baselines. 
Finally, \method outperforms existing prompt optimization methods across all settings, further demonstrating the effectiveness of our strategies and adaptive policy in optimizing prompts solely based on preference feedback.
In summary, these findings validate the design of \method, the effectiveness of individual strategies, and the overall robustness.

%% file: sections/x_appendix.tex
\section{Meta-Prompts used in \method }

We present the meta-prompts used in \method for optimizing prompts in an image-generation task.
These meta-prompts can be easily adapted to other generation tasks by simply rephrasing the task description.

\subsection{Random Exploration}

In the first iteration, because no historical information about the user's preferences is available, \method generates prompts to explore diverse directions using the following meta-prompt:

\begin{lstlisting}
Write {variant_num} variants based on this prompt
{initial_prompt}

Each variant should:
- Remain the objects and their composition in the old prompt
- Explore different style directions, as much different as possible.
- Be concise (under 60 words)
- Include visually rich and specific details
- Be diverse in composition and wording

Output only the {variant_num} new prompts. Each on its own line. No explanation.
\end{lstlisting}

\subsection{Alignment}

In the \emph{Alignment} strategy, \method first summarizes the prompts that users encountered in previous iterations:

\begin{lstlisting}
You are an assistant that extracts **stylistic and aesthetic trends** from prompt examples.

Below are previously preferred prompts:
{preferred_history}

Below are previously unpreferred prompts:
{unpreferred_history}

Your task:

1. Identify the **distinct visual styles or aesthetics** (e.g., cyberpunk, golden hour, anime, photorealism, minimalism, surrealism, vintage, etc.) that appear in the preferred prompts. Be specific and name the styles.
2. Identify which **specific styles or characteristics** appear frequently in the unpreferred prompts.
3. Clearly conclude which styles are **consistently preferred** and which are **not preferred**, based on these examples.
4. Summarize this as:
- A short paragraph describing **which styles are most favored**, with examples of good visual traits (e.g., dramatic lighting, deep contrast, cinematic framing).
- A short paragraph describing **which styles or traits are less favored or problematic**, with examples (e.g., flat lighting, generic scenery, overused tropes).
5. Use clear, **style-labeled language** and avoid repeating full prompt texts.

Format:
Preferred style summary:
...

Unpreferred style summary:
...

Style preference conclusion:
...
\end{lstlisting}

Then, \method integrates the history summary with the user's selected prompts from the current iteration and generates the textual gradient as follows:

\begin{lstlisting}
You are assisting in refining image generation prompts based on user preferences.

Users have shown a clear preference for the following prompts:
{preferred_prompts}

In contrast, the following prompts were not preferred:
{non_preferred_prompts}

Summary of preferences from earlier iterations:
{summary}

Please analyze the visual differences between the preferred and unpreferred image prompts, focusing especially on the **stylistic features and fine-grained visual aesthetics** that each prompt produces.

Identify what **specific styles** (e.g., cinematic, minimalist, painterly, photorealistic, surreal, vintage, anime, etc.) or **visual characteristics** (e.g., lighting, texture, composition, color grading, camera angle, subject positioning, background complexity) are preferred.

Compare these features against those in the unpreferred prompts, and describe what **key visual elements or stylistic patterns** are lacking or less desirable.

Provide a clear, **style-focused** piece of feedback that reveals how future prompts can better align with the preferred visual outcomes.

Respond only with the feedback, as it will be used as a global gradient signal to enhance prompt quality.
\end{lstlisting}

With the textual gradient, \method aligns the non-preferred prompts using the following meta-prompts:

\begin{lstlisting}
You are an expert prompt engineer for text-to-image generation. Your task is to rewrite and improve the original prompt to create more preferred image outputs.

The following prompts were not preferred by users:
{non_preferred_prompts}

Summarized feedback on why these prompts may be suboptimal:
{gradients}

Using this feedback, generate {num_variants - len(preferred_prompts)} improved prompt variants that:
- Retain all key objects and their arrangement from the original prompt: {initial_prompt}
- Draw inspiration from the original unpreferred prompts
- Address the common issues highlighted in the feedback
- Are concise (under 60 words)
- Include vivid, specific, and visually rich details
- Use feedback constructively, but do not overfit, allow creative detours

Output exactly {num_variants - len(preferred_prompts)} revised prompts, each on its own line, with no explanation.
\end{lstlisting}

\subsection{Expansion}

In the Expansion strategy, \method applies an evolutionary algorithm (EA) to evolve the prompts.
It first uses the user's preferred prompts from the current iteration as the population for performing crossover with the following prompts:

\begin{lstlisting}
You are an AI assistant simulating crossover for evolutionary prompt optimization in a text-to-image generation task.

Each parent prompt describes the **same object or scene**, but in a **different visual style** (e.g., art medium, color palette, mood, lighting, texture, rendering technique, etc.).

The goal is to generate **three child prompts** that:
1. **Preserve the object or scene** described in both parent prompts
2. **Blend, recombine, or hybridize the stylistic elements** of the parents
3. Explore **diverse combinations** of style modifiers (without assuming a fixed target style)
4. Avoid exact repetition of full parent prompts
5. Can serve as exploratory candidates for downstream evaluation (e.g., human preference, image similarity, aesthetic score)
6. Each child prompt is under 60 words.

Be creative and descriptive. Focus on **visual and stylistic traits** such as mood, lighting, rendering, detail level, color scheme, texture, medium, or technique. You may keep or alter the style names (e.g., "steampunk", "cyberpunk", "oil painting", etc.), or omit them entirely in favor of descriptive traits.

---

Example Input:

Parent A: "a futuristic city skyline at dusk, rendered in low-poly 3D style with pastel colors"  
Parent B: "a futuristic city skyline at dusk, digital painting with painterly brushstrokes and glowing lights"

Output:
Child 1: "a futuristic city skyline at dusk, a fusion of low-poly geometry and painterly textures, glowing softly in pastel tones"  
Child 2: "a futuristic city skyline at dusk, rendered in semi-abstract style with blocky shapes and expressive lighting"  
Child 3: "a futuristic city skyline at dusk, with glowing brushstrokes, soft gradients, and minimalist polygonal forms"

---

Now, process the following two parents and output 1 new, diverse child prompts:
Prompt 1: {prompt1}
Prompt 2: {prompt2}

Only return the child prompt.
\end{lstlisting}

Then, we perform mutations on each child prompt:

\begin{lstlisting}
You are an AI assistant simulating **mutation** in an evolutionary algorithm for optimizing text-to-image prompts.

---

## Task
Given **one input prompt** describing a specific object or scene, generate **{child_prompt_num} mutated versions** by changing its **stylistic elements** while keeping the **core subject** exactly the same.

---

## Mutation Rules

### 1. Preserve the Subject
- The main object or scene described in the original prompt must remain unchanged.
- Do **not** add, remove, or replace the main subject or setting.

### 2. Modify Style-Related Elements
Make changes only to style-oriented descriptors, such as:
- **Color palette** (e.g., warm earthy tones -> neon pastels)
- **Lighting** (e.g., soft morning light -> dramatic chiaroscuro)
- **Rendering technique** (e.g., watercolor -> photorealism)
- **Medium** (e.g., pencil sketch -> digital painting)
- **Mood/atmosphere** (e.g., serene -> chaotic)
- **Texture** (e.g., smooth glassy -> rough, grainy)
- **Level of abstraction** (e.g., hyperrealistic -> minimalistic)
- **Other visual descriptors** relevant to style

### 3. Control the Degree of Change with `intensity`
The `intensity` parameter determines **how far** the mutation deviates from the original style:
- **intensity = 0** -> Minimal stylistic change  
- Alter **only one** style element slightly.  
- Keep most wording and details identical.  
- Small, subtle variations.  

- **intensity = 1** -> Maximum stylistic change (**totally random variation**)  
- Change **as many style elements as possible** while keeping the subject recognizable.  
- Explore **completely different artistic styles, lighting, moods, mediums, and composition choices**.  
- Use **different vocabulary, sentence structure, and tone** for each output.  
- Every mutated prompt should feel **radically different** in style from both the original and from each other.

---

## Examples

**Example Input Prompt:**  
"A medieval castle on a hill at sunrise, painted in watercolor with soft pastel colors."

**intensity = 0** (minimal change examples):  
1. "A medieval castle on a hill at sunrise, painted in watercolor with slightly warmer golden tones."  
2. "A medieval castle on a hill at sunrise, painted in watercolor with a cooler, misty color palette."

**intensity = 1** (totally random style examples):  
1. "In a futuristic neon cityscape, a towering medieval castle rises above the skyline, glowing in electric blues and magentas, rendered in glitch-art style."  
2. "A dreamlike medieval castle perched high on a hill, sculpted entirely from molten glass, catching the light in fractal rainbows, in ultra-detailed 8K realism."  
3. "An abstract cubist interpretation of a medieval castle on a hill, broken into geometric shards of crimson, gold, and midnight blue."  
4. "A whimsical claymation scene of a medieval castle, its turrets swaying slightly under a pink cotton-candy sky."  
5. "A dark, cinematic shot of a medieval castle during a raging storm, lit only by flashes of lightning, rendered in gritty black-and-white film grain."

---

## Output Format
- Generate exactly **{child_prompt_num}** mutated prompts.
- Each prompt should be on its own line.
- Do **not** include numbering, bullets, or explanations.
- Under 60 words.

---

Current mutation intensity: {mutate_intensity}

**Input Prompt to Mutate:**  
{prompt}
\end{lstlisting}

\subsection{Consistency Check}

At the final stage, \method checks the optimized prompts to ensure that all essential information from the initial prompt is preserved, and restores any missing information.

\begin{lstlisting}
You are a helpful assistant for prompt evaluation and refinement.

Your task is to:
1. Reflect on whether the following revised prompt retains all important objects, relationships, and visual elements from the initial prompt.
2. If anything important is missing, revise the prompt to add those missing elements while preserving clarity, conciseness (under 60 words), and visual richness.
3. If nothing is missing, return the original revised prompt as-is.
4. Keep the revised prompt under 60 words.

Initial prompt:
"{initial_prompt}"

Revised prompt:
"{prompt}"

Only respond with the final prompt (either unchanged or improved to restore missing content)
\end{lstlisting}

\subsection{Generating Target Prompt for Synthetic Tests}
\label{sec:ablation_prompt}

We use the following prompt to generate the \textit{low-granularity} target prompts in our synthetic tests.

\begin{lstlisting}
Extend the following prompt by appending a short artistic or stylistic vibe (e.g., "surreal", "dreamlike", "sci-fi", "vintage") that complements the visual tone and theme of the original prompt.

- Use only 1 to 3 words.
- The added phrase should enhance or harmonize with the mood and imagery already present.
- Avoid disrupting the original meaning or introducing new objects.

Prompt:
{initial_prompt}
Only return the full modified prompt with the added style at the end.
\end{lstlisting}

We use the following prompt to generate the \textit{high-granularity} target prompts.

\begin{lstlisting}
Enhance the following prompt by adding descriptive details to each object.

Prompt:
{initial_prompt}
Ensure all original objects remain in the revised prompt. The enhanced prompt must be a single line and under 60 words.

Return only the modified prompt.
\end{lstlisting}

We use the following prompt to generate the target prompt for creative writing.

\begin{lstlisting}
Rewrite the following writing prompt to make it more engaging by adding a specific story style (such as sci-fi, fantasy, mystery, horror, or any creative genre) chosen by you.
Keep it short, clear, and suitable as a story prompt:
{prompt}
The new prompt should have a similar length to the original one.
Only return the new prompt. No explanation. No reasoning.
\end{lstlisting}

\section{Full statistical analysis results}
\label{sec:full_statistical_results}

We here present the full statistical analysis results of our user study.

\begin{table*}[h]
\centering
\caption{Statistical results for all metrics. Abbreviations: (a: \emph{\method}, c: \emph{Clarification}, d: \emph{DSPy}, p: \emph{PromptCharm}). (*: $p < 0.05$, **: $p < 0.01$, ***: $p < 0.001$).}
\label{tab:user_study_full_results}
\Description{The table provides statistical results for different metrics across two task types: Close-ended and Open-ended. The metrics are grouped into three categories: efficiency metrics (Iteration and Total Time), NASA-TLX workload metrics (Mental Demand, Physical Demand, Temporal Demand, Performance, Effort, and Frustration), and CSI user experience metrics (Enjoyment, Exploration, Expressiveness, Immersion, and Results Worth Effort). Highly significant differences ($p<0.001$, denoted by ***) are observed across all four conditions (APPO, Clarification, DSPy, and PromptCharm) for the primary efficiency metrics (Iteration and Total Time) and the majority of workload metrics (Mental Demand, Physical Demand, Performance, Effort, and Frustration) in both task types, with Temporal Demand showing no significant difference. Similarly, for user experience, significant differences are found in Exploration, Expressiveness, and Results Worth Effort in the open-ended task, while close-ended tasks only show significance for Exploration, Expressiveness, and Results Worth Effort, with Immersion and Enjoyment showing no significant differences in either task type.}
\begin{tabular}{c|c|c}
\hline
Metrics & Close-ended & Open-ended \\ 
\hline
Iteration & \anova{2}{30}{18.33}{<}{0.001}{}{***} & \anova{2}{30}{14.05}{<}{0.001}{}{***} \\
Total time & \anova{3}{45}{101.98}{<}{0.001}{}{***} & \anova{3}{45}{114.72}{<}{0.001}{}{***} \\ 
\hline
Mental Demand & \friedman{3}{22.85}{<}{0.001}{***} & \friedman{3}{24.82}{<}{0.001}{***} \\
Physical Demand & \friedman{3}{36.02}{<}{0.001}{***} & \friedman{3}{36.49}{<}{0.001}{***} \\
Temporal Demand & \friedman{3}{5.52}{>}{0.05} & \friedman{3}{4.15}{>}{0.05} \\
Performance & \friedman{3}{15.46}{<}{0.01}{**} & \friedman{3}{27.92}{<}{0.001}{***} \\
Effort & \friedman{3}{38.73}{<}{0.001}{***} & \friedman{3}{43.46}{<}{0.001}{***} \\
Frustration & \friedman{3}{32.76}{<}{0.001}{***} & \friedman{3}{30.71}{<}{0.001}{***} \\
\hline
Enjoyment & \friedman{3}{4.59}{>}{0.05} & \friedman{3}{3.75}{>}{0.05} \\
Exploration & \friedman{3}{7.87}{<}{0.05}{*} & \friedman{3}{22.70}{<}{0.001}{***} \\
Expressiveness & \friedman{3}{20.01}{<}{0.001}{***} & \friedman{3}{33.02}{<}{0.001}{***} \\
Immersion & \friedman{3}{7.37}{>}{0.05} & \friedman{3}{1.84}{>}{0.05} \\
Results Worth Effort & \friedman{3}{9.23}{<}{0.05}{*} & \friedman{3}{9.46}{<}{0.05}{*} \\
\hline
\end{tabular}
\end{table*}

\begin{table*}[h]
\centering
\caption{
Post hoc analysis for CSI metrics.
Only significant pairwise differences are reported.
Abbreviations: (a: \emph{\method}, c: \emph{Clarification}, d: \emph{DSPy}, p: \emph{PromptCharm}). 
(*: $p < 0.05$, **: $p < 0.01$, ***: $p < 0.001$).
}
\label{tab:user_study_full_post_hoc_csi_results}
\Description{The table presents the post-hoc analysis for the CSI metrics (Enjoyment, Exploration, Expressiveness, Immersion, and Results Worth Effort) across four conditions (APPO (a), Clarification (c), DSPy (d), and PromptCharm (p)) for Close-ended and Open-ended tasks. For Close-ended tasks, Enjoyment and Immersion show no significant differences between any pairs. However, Expressiveness reveals APPO is significantly different from DSPy ($p<0.01$) and PromptCharm ($p<0.01$), and Clarification is significantly different from DSPy ($p<0.05$) and PromptCharm ($p<0.05$). For Open-ended tasks, significant differences are concentrated in Exploration and Expressiveness, where APPO is significantly different from all other methods in Exploration ($p<0.05$), and in Expressiveness, APPO is significantly different from DSPy ($p<0.001$) and PromptCharm ($p<0.01$), and Clarification is significantly different from DSPy ($p<0.01$). Results Worth Effort also shows several significant differences in both task types, notably APPO differing from PromptCharm in both categories ($p<0.05$).
}
\begin{tabular}{c|l|l}
\hline
\textbf{Metrics} & \textbf{Close-ended} & \textbf{Open-ended} \\
\hline

\multirow{7}{*}{Enjoyment}
 & a--c: \wilcoxon{11.00}{>}{0.05}{} & a--c: \wilcoxon{34.50}{>}{0.05}{} \\
 & a--d: \wilcoxon{31.00}{>}{0.05}{} & a--d: \wilcoxon{16.50}{>}{0.05}{} \\
 & a--p: \wilcoxon{21.50}{>}{0.05}{} & a--p: \wilcoxon{42.50}{>}{0.05}{} \\
 & c--d: \wilcoxon{28.50}{>}{0.05}{} & c--d: \wilcoxon{35.00}{>}{0.05}{} \\
 & c--p: \wilcoxon{25.50}{>}{0.05}{} & c--p: \wilcoxon{10.00}{>}{0.05}{} \\
 & d--p: \wilcoxon{29.50}{>}{0.05}{} & d--p: \wilcoxon{43.00}{>}{0.05}{} \\
\hline

\multirow{7}{*}{Exploration}
 & a--c: \wilcoxon{3.00}{<}{0.05}{*} & a--c: \wilcoxon{8.00}{<}{0.05}{*} \\
 & a--d: \wilcoxon{12.00}{>}{0.05}{} & a--d: \wilcoxon{0.00}{<}{0.05}{*} \\
 & a--p: \wilcoxon{15.50}{>}{0.05}{} & a--p: \wilcoxon{14.00}{<}{0.05}{*} \\
 & c--d: \wilcoxon{3.00}{>}{0.05}{} & c--d: \wilcoxon{6.50}{<}{0.05}{*} \\
 & c--p: \wilcoxon{15.00}{>}{0.05}{} & c--p: \wilcoxon{7.00}{>}{0.05}{} \\
 & d--p: \wilcoxon{13.50}{>}{0.05}{} & d--p: \wilcoxon{15.00}{<}{0.05}{*} \\
\hline

\multirow{7}{*}{Expressiveness}
 & a--c: \wilcoxon{17.00}{>}{0.05}{} & a--c: \wilcoxon{5.00}{<}{0.05}{*} \\
 & a--d: \wilcoxon{4.00}{<}{0.01}{**} & a--d: \wilcoxon{0.00}{<}{0.001}{***} \\
 & a--p: \wilcoxon{0.00}{<}{0.01}{**} & a--p: \wilcoxon{0.00}{<}{0.01}{**} \\
 & c--d: \wilcoxon{5.00}{<}{0.05}{*} & c--d: \wilcoxon{0.00}{<}{0.01}{**} \\
 & c--p: \wilcoxon{0.00}{<}{0.05}{*} & c--p: \wilcoxon{12.00}{>}{0.05}{} \\
 & d--p: \wilcoxon{13.50}{>}{0.05}{} & d--p: \wilcoxon{13.50}{>}{0.05}{} \\
\hline

\multirow{7}{*}{Immersion}
 & a--c: \wilcoxon{13.00}{>}{0.05}{} & a--c: \wilcoxon{20.00}{>}{0.05}{} \\
 & a--d: \wilcoxon{15.00}{>}{0.05}{} & a--d: \wilcoxon{33.50}{>}{0.05}{} \\
 & a--p: \wilcoxon{10.50}{>}{0.05}{} & a--p: \wilcoxon{27.50}{>}{0.05}{} \\
 & c--d: \wilcoxon{2.00}{>}{0.05}{} & c--d: \wilcoxon{15.00}{>}{0.05}{} \\
 & c--p: \wilcoxon{9.00}{>}{0.05}{} & c--p: \wilcoxon{24.00}{>}{0.05}{} \\
 & d--p: \wilcoxon{12.00}{>}{0.05}{} & d--p: \wilcoxon{13.50}{>}{0.05}{} \\
\hline

\multirow{7}{*}{Results Worth Effort}
 & a--c: \wilcoxon{20.00}{>}{0.05}{} & a--c: \wilcoxon{1.00}{<}{0.05}{*} \\
 & a--d: \wilcoxon{25.50}{>}{0.05}{} & a--d: \wilcoxon{6.50}{<}{0.05}{*} \\
 & a--p: \wilcoxon{5.00}{<}{0.05}{*} & a--p: \wilcoxon{3.50}{<}{0.05}{*} \\
 & c--d: \wilcoxon{33.50}{>}{0.05}{} & c--d: \wilcoxon{22.50}{>}{0.05}{} \\
 & c--p: \wilcoxon{3.50}{<}{0.05}{*} & c--p: \wilcoxon{32.50}{>}{0.05}{} \\
 & d--p: \wilcoxon{3.50}{<}{0.05}{*} & d--p: \wilcoxon{9.00}{>}{0.05}{} \\
\hline

\end{tabular}
\end{table*}

\begin{table*}[h]
\centering
\caption{
Post hoc analysis for iteration, total time, and NASA-TLX metrics.
Only significant pairwise differences are reported.
Abbreviations: (a: \emph{\method}, c: \emph{Clarification}, d: \emph{DSPy}, p: \emph{PromptCharm}). 
(*: $p < 0.05$, **: $p < 0.01$, ***: $p < 0.001$).
}
\label{tab:user_study_full_post_hoc_results}
\Description{The table presents the post-hoc analysis for efficiency (Iteration, Total Time) and NASA-TLX workload metrics across four conditions: APPO (a), Clarification (c), DSPy (d), and PromptCharm (p), for both Close-ended and Open-ended tasks. For Iteration, APPO and DSPy show no significant difference from each other but are significantly better than PromptCharm in both task types, with DSPy being highly superior to PromptCharm in close-ended tasks ($p<0.001$). For Total Time, all conditions are significantly different from each other ($p<0.001$) except for the pairs Clarification versus DSPy, indicating a widespread difference in time efficiency. Across the NASA-TLX metrics, Temporal Demand shows no significant pairwise differences in either task type. Conversely, APPO is significantly different from all other methods for Mental Demand, Physical Demand, Effort, and Frustration in the Close-ended task (most at $p<0.001$), generally indicating lower perceived workload for APPO. For Open-ended tasks, APPO maintains a high number of significant differences, particularly showing a highly significant difference from all three other methods for Effort ($p<0.001$).}
\begin{tabular}{c|l|l}
\hline
\textbf{Metrics} & \textbf{Close-ended} & \textbf{Open-ended} \\
\hline

\multirow{3}{*}{Iteration}
 & a--d: \ttest{2}{0.38}{>}{0.05}{} & a--d: \ttest{2}{0.60}{>}{0.05}{} \\
 & a--p: \ttest{2}{4.42}{<}{0.01}{**} & a--p: \ttest{2}{4.19}{<}{0.01}{**} \\
 & d--p: \ttest{2}{5.37}{<}{0.001}{***} & d--p: \ttest{2}{3.82}{<}{0.05}{*} \\
\hline

\multirow{7}{*}{Total time}
 & a--c: \ttest{2}{4.74}{<}{0.001}{***} & a--c: \ttest{2}{8.90}{<}{0.001}{***} \\
 & a--d: \ttest{2}{6.02}{<}{0.001}{***} & a--d: \ttest{2}{7.00}{<}{0.001}{***} \\
 & a--p: \ttest{2}{8.55}{<}{0.001}{***} & a--p: \ttest{2}{9.31}{<}{0.001}{***} \\
 & c--d: \ttest{2}{-1.88}{>}{0.05}{} & c--d: \ttest{2}{-1.73}{>}{0.05}{} \\
 & c--p: \ttest{2}{6.58}{<}{0.001}{***} & c--p: \ttest{2}{6.86}{<}{0.001}{***} \\
 & d--p: \ttest{2}{10.37}{<}{0.001}{***} & d--p: \ttest{2}{10.45}{<}{0.001}{***} \\
\hline

\multirow{7}{*}{Mental Demand}
 & a--c: \wilcoxon{1.00}{<}{0.001}{***} & a--c: \wilcoxon{11.00}{<}{0.05}{*} \\
 & a--d: \wilcoxon{0.00}{<}{0.001}{***} & a--d: \wilcoxon{1.00}{<}{0.01}{**} \\
 & a--p: \wilcoxon{1.50}{<}{0.001}{***} & a--p: \wilcoxon{0.00}{<}{0.01}{**} \\
 & c--d: \wilcoxon{46.50}{>}{0.05}{} & c--d: \wilcoxon{21.50}{>}{0.05}{} \\
 & c--p: \wilcoxon{49.00}{>}{0.05}{} & c--p: \wilcoxon{8.00}{<}{0.05}{*} \\
 & d--p: \wilcoxon{61.50}{>}{0.05}{} & d--p: \wilcoxon{54.00}{>}{0.05}{} \\
\hline

\multirow{7}{*}{Physical Demand}
 & a--c: \wilcoxon{1.00}{<}{0.01}{**} & a--c: \wilcoxon{1.50}{<}{0.01}{**} \\
 & a--d: \wilcoxon{0.00}{<}{0.01}{**} & a--d: \wilcoxon{0.00}{<}{0.01}{**} \\
 & a--p: \wilcoxon{0.00}{<}{0.01}{**} & a--p: \wilcoxon{0.00}{<}{0.01}{**} \\
 & c--d: \wilcoxon{22.50}{>}{0.05}{} & c--d: \wilcoxon{4.00}{<}{0.01}{**} \\
 & c--p: \wilcoxon{45.00}{>}{0.05}{} & c--p: \wilcoxon{0.00}{<}{0.01}{**} \\
 & d--p: \wilcoxon{11.50}{>}{0.05}{} & d--p: \wilcoxon{15.00}{>}{0.05}{} \\
\hline

\multirow{7}{*}{Temporal Demand}
 & a--c: \wilcoxon{13.00}{>}{0.05}{} & a--c: \wilcoxon{21.00}{>}{0.05}{} \\
 & a--d: \wilcoxon{28.50}{>}{0.05}{} & a--d: \wilcoxon{24.50}{>}{0.05}{} \\
 & a--p: \wilcoxon{23.50}{>}{0.05}{} & a--p: \wilcoxon{20.00}{>}{0.05}{} \\
 & c--d: \wilcoxon{27.00}{>}{0.05}{} & c--d: \wilcoxon{25.50}{>}{0.05}{} \\
 & c--p: \wilcoxon{46.50}{>}{0.05}{} & c--p: \wilcoxon{26.00}{>}{0.05}{} \\
 & d--p: \wilcoxon{9.00}{>}{0.05}{} & d--p: \wilcoxon{41.00}{>}{0.05}{} \\
\hline

\multirow{7}{*}{Performance}
 & a--c: \wilcoxon{10.00}{<}{0.05}{*} & a--c: \wilcoxon{12.50}{<}{0.05}{*} \\
 & a--d: \wilcoxon{53.00}{>}{0.05}{} & a--d: \wilcoxon{17.00}{>}{0.05}{} \\
 & a--p: \wilcoxon{9.50}{<}{0.05}{*} & a--p: \wilcoxon{11.50}{<}{0.01}{**} \\
 & c--d: \wilcoxon{3.50}{<}{0.05}{*} & c--d: \wilcoxon{0.00}{<}{0.01}{**} \\
 & c--p: \wilcoxon{41.50}{>}{0.05}{} & c--p: \wilcoxon{18.00}{>}{0.05}{} \\
 & d--p: \wilcoxon{15.50}{>}{0.05}{} & d--p: \wilcoxon{0.00}{<}{0.01}{**} \\
\hline

\multirow{7}{*}{Effort}
 & a--c: \wilcoxon{0.00}{<}{0.001}{***} & a--c: \wilcoxon{0.00}{<}{0.001}{***} \\
 & a--d: \wilcoxon{0.00}{<}{0.001}{***} & a--d: \wilcoxon{0.00}{<}{0.001}{***} \\
 & a--p: \wilcoxon{0.00}{<}{0.001}{***} & a--p: \wilcoxon{0.00}{<}{0.001}{***} \\
 & c--d: \wilcoxon{16.00}{<}{0.05}{*} & c--d: \wilcoxon{0.00}{<}{0.01}{**} \\
 & c--p: \wilcoxon{20.00}{>}{0.05}{} & c--p: \wilcoxon{15.00}{>}{0.05}{} \\
 & d--p: \wilcoxon{31.50}{>}{0.05}{} & d--p: \wilcoxon{2.00}{<}{0.01}{**} \\
\hline

\multirow{7}{*}{Frustration}
 & a--c: \wilcoxon{0.00}{<}{0.001}{***} & a--c: \wilcoxon{1.00}{<}{0.01}{**} \\
 & a--d: \wilcoxon{0.00}{<}{0.01}{***} & a--d: \wilcoxon{0.00}{<}{0.01}{**} \\
 & a--p: \wilcoxon{0.00}{<}{0.01}{**} & a--p: \wilcoxon{0.00}{<}{0.01}{**} \\
 & c--d: \wilcoxon{26.50}{>}{0.05}{} & c--d: \wilcoxon{3.00}{<}{0.05}{*} \\
 & c--p: \wilcoxon{7.00}{<}{0.05}{*} & c--p: \wilcoxon{31.50}{>}{0.05}{} \\
 & d--p: \wilcoxon{28.00}{>}{0.05}{} & d--p: \wilcoxon{5.00}{>}{0.05}{} \\
\hline

\end{tabular}
\end{table*}

%% file: sample-base.bib
@inproceedings{brade2023promptify,
  title={Promptify: Text-to-image generation through interactive prompt exploration with large language models},
  author={Brade, Stephen and Wang, Bryan and Sousa, Mauricio and Oore, Sageev and Grossman, Tovi},
  booktitle={Proceedings of the 36th Annual ACM Symposium on User Interface Software and Technology},
  pages={1--14},
  year={2023}
}

@inproceedings{wang2024promptcharm,
  title={Promptcharm: Text-to-image generation through multi-modal prompting and refinement},
  author={Wang, Zhijie and Huang, Yuheng and Song, Da and Ma, Lei and Zhang, Tianyi},
  booktitle={Proceedings of the 2024 CHI Conference on Human Factors in Computing Systems},
  pages={1--21},
  year={2024}
}

@inproceedings{chung2023promptpaint,
  title={Promptpaint: Steering text-to-image generation through paint medium-like interactions},
  author={Chung, John Joon Young and Adar, Eytan},
  booktitle={Proceedings of the 36th Annual ACM Symposium on User Interface Software and Technology},
  pages={1--17},
  year={2023}
}

@article{guo2024prompthis,
  title={Prompthis: Visualizing the process and influence of prompt editing during text-to-image creation},
  author={Guo, Yuhan and Shao, Hanning and Liu, Can and Xu, Kai and Yuan, Xiaoru},
  journal={IEEE Transactions on Visualization and Computer Graphics},
  year={2024},
  publisher={IEEE}
}

@inproceedings{peng2024designprompt,
  title={Designprompt: Using multimodal interaction for design exploration with generative ai},
  author={Peng, Xiaohan and Koch, Janin and Mackay, Wendy E},
  booktitle={Proceedings of the 2024 ACM Designing Interactive Systems Conference},
  pages={804--818},
  year={2024}
}

@inproceedings{10.1145/3706598.3713603,
author = {Liao, Yi-Chi and Streli, Paul and Li, Zhipeng and Gebhardt, Christoph and Holz, Christian},
title = {Continual Human-in-the-Loop Optimization},
year = {2025},
isbn = {9798400713941},
publisher = {Association for Computing Machinery},
address = {New York, NY, USA},
url = {https://doi.org/10.1145/3706598.3713603},
doi = {10.1145/3706598.3713603},
booktitle = {Proceedings of the 2025 CHI Conference on Human Factors in Computing Systems},
articleno = {795},
numpages = {26},
location = {
},
series = {CHI '25}
}

@inproceedings{Liao2026HOMI,
  author    = {Liao, Yi-Chi and Belo, Jo{\~a}o and Moon, Hee-Seung and Steimle, J{\"u}rgen and Feit, Anna Maria},
  title     = {Efficient Human-in-the-Loop Optimization via Priors Learned from User Models},
  booktitle = {Proceedings of the 2026 {CHI} Conference on Human Factors in Computing Systems},
  series    = {CHI '26},
  year      = {2026},
  month     = apr,
  address   = {Barcelona, Spain},
  publisher = {ACM},
  location  = {Barcelona, Spain},
  numpages  = {24},
  doi       = {10.1145/3772318.3791976},
  url       = {https://doi.org/10.1145/3772318.3791976}
}

@inproceedings{masson2024directgpt,
  title={Directgpt: A direct manipulation interface to interact with large language models},
  author={Masson, Damien and Malacria, Sylvain and Casiez, G{\'e}ry and Vogel, Daniel},
  booktitle={Proceedings of the 2024 CHI Conference on Human Factors in Computing Systems},
  pages={1--16},
  year={2024}
}

@inproceedings{arawjo2024chainforge,
  title={Chainforge: A visual toolkit for prompt engineering and llm hypothesis testing},
  author={Arawjo, Ian and Swoopes, Chelse and Vaithilingam, Priyan and Wattenberg, Martin and Glassman, Elena L},
  booktitle={Proceedings of the 2024 CHI Conference on Human Factors in Computing Systems},
  pages={1--18},
  year={2024}
}

@article{feng2023promptmagician,
  title={Promptmagician: Interactive prompt engineering for text-to-image creation},
  author={Feng, Yingchaojie and Wang, Xingbo and Wong, Kam Kwai and Wang, Sijia and Lu, Yuhong and Zhu, Minfeng and Wang, Baicheng and Chen, Wei},
  journal={IEEE Transactions on Visualization and Computer Graphics},
  volume={30},
  number={1},
  pages={295--305},
  year={2023},
  publisher={IEEE}
}

@article{fernando2023promptbreeder,
  title={Promptbreeder: Self-referential self-improvement via prompt evolution},
  author={Fernando, Chrisantha and Banarse, Dylan and Michalewski, Henryk and Osindero, Simon and Rockt{\"a}schel, Tim},
  journal={arXiv preprint arXiv:2309.16797},
  year={2023}
}

@article{pryzant2023automatic,
  title={Automatic prompt optimization with" gradient descent" and beam search},
  author={Pryzant, Reid and Iter, Dan and Li, Jerry and Lee, Yin Tat and Zhu, Chenguang and Zeng, Michael},
  journal={arXiv preprint arXiv:2305.03495},
  year={2023}
}

@article{lin2024prompt,
  title={Prompt optimization with human feedback},
  author={Lin, Xiaoqiang and Dai, Zhongxiang and Verma, Arun and Ng, See-Kiong and Jaillet, Patrick and Low, Bryan Kian Hsiang},
  journal={arXiv preprint arXiv:2405.17346},
  year={2024}
}

@article{dong2023pace,
  title={Pace: Improving prompt with actor-critic editing for large language model},
  author={Dong, Yihong and Luo, Kangcheng and Jiang, Xue and Jin, Zhi and Li, Ge},
  journal={arXiv preprint arXiv:2308.10088},
  year={2023}
}

@article{hao2023optimizing,
  title={Optimizing prompts for text-to-image generation},
  author={Hao, Yaru and Chi, Zewen and Dong, Li and Wei, Furu},
  journal={Advances in Neural Information Processing Systems},
  volume={36},
  pages={66923--66939},
  year={2023}
}

@inproceedings{mo2024dynamic,
  title={Dynamic prompt optimizing for text-to-image generation},
  author={Mo, Wenyi and Zhang, Tianyu and Bai, Yalong and Su, Bing and Wen, Ji-Rong and Yang, Qing},
  booktitle={Proceedings of the IEEE/CVF Conference on Computer Vision and Pattern Recognition},
  pages={26627--26636},
  year={2024}
}

@article{manas2024improving,
  title={Improving text-to-image consistency via automatic prompt optimization},
  author={Ma{\~n}as, Oscar and Astolfi, Pietro and Hall, Melissa and Ross, Candace and Urbanek, Jack and Williams, Adina and Agrawal, Aishwarya and Romero-Soriano, Adriana and Drozdzal, Michal},
  journal={arXiv preprint arXiv:2403.17804},
  year={2024}
}

@article{wen2023hard,
  title={Hard prompts made easy: Gradient-based discrete optimization for prompt tuning and discovery},
  author={Wen, Yuxin and Jain, Neel and Kirchenbauer, John and Goldblum, Micah and Geiping, Jonas and Goldstein, Tom},
  journal={Advances in Neural Information Processing Systems},
  volume={36},
  pages={51008--51025},
  year={2023}
}

@inproceedings{hu2023tifa,
  title={Tifa: Accurate and interpretable text-to-image faithfulness evaluation with question answering},
  author={Hu, Yushi and Liu, Benlin and Kasai, Jungo and Wang, Yizhong and Ostendorf, Mari and Krishna, Ranjay and Smith, Noah A},
  booktitle={Proceedings of the IEEE/CVF International Conference on Computer Vision},
  pages={20406--20417},
  year={2023}
}

@inproceedings{mahdavi2024ai,
  title={Is it AI or is it me? Understanding users’ prompt journey with text-to-image generative AI tools},
  author={Mahdavi Goloujeh, Atefeh and Sullivan, Anne and Magerko, Brian},
  booktitle={Proceedings of the 2024 CHI Conference on Human Factors in Computing Systems},
  pages={1--13},
  year={2024}
}

@article{rose2010automatic,
  title={Automatic keyword extraction from individual documents},
  author={Rose, Stuart and Engel, Dave and Cramer, Nick and Cowley, Wendy},
  journal={Text mining: applications and theory},
  pages={1--20},
  year={2010},
  publisher={Wiley Online Library}
}

@article{mishra2023promptaid,
  title={Promptaid: Prompt exploration, perturbation, testing and iteration using visual analytics for large language models},
  author={Mishra, Aditi and Soni, Utkarsh and Arunkumar, Anjana and Huang, Jinbin and Kwon, Bum Chul and Bryan, Chris},
  journal={arXiv preprint arXiv:2304.01964},
  year={2023}
}

@inproceedings{drosos2025dynamic,
  title={Dynamic Prompt Middleware: Contextual Prompt Refinement Controls for Comprehension Tasks},
  author={Drosos, Ian and Williams, Jack and Sarkar, Advait and Wilson, Nicholas and Rintel, Sean and Panda, Payod},
  booktitle={Proceedings of the 4th Annual Symposium on Human-Computer Interaction for Work},
  pages={1--23},
  year={2025}
}

@inproceedings{peng2025fusain,
  title={FusAIn: Composing Generative AI Visual Prompts Using Pen-based Interaction},
  author={Peng, Xiaohan and Koch, Janin and Mackay, Wendy E},
  booktitle={Proceedings of the 2025 CHI Conference on Human Factors in Computing Systems},
  pages={1--20},
  year={2025}
}

@article{xiang2025sel3dcraft,
  title={Sel3DCraft: Interactive Visual Prompts for User-Friendly Text-to-3D Generation},
  author={Xiang, Nan and Liang, Tianyi and Huang, Haiwen and Jiang, Shiqi and Huang, Hao and Huang, Yifei and Chen, Liangyu and Wang, Changbo and Li, Chenhui},
  journal={arXiv preprint arXiv:2508.00428},
  year={2025}
}

@article{wang2024discrete,
  title={On discrete prompt optimization for diffusion models},
  author={Wang, Ruochen and Liu, Ting and Hsieh, Cho-Jui and Gong, Boqing},
  journal={arXiv preprint arXiv:2407.01606},
  year={2024}
}

@inproceedings{yang2023large,
  title={Large language models as optimizers},
  author={Yang, Chengrun and Wang, Xuezhi and Lu, Yifeng and Liu, Hanxiao and Le, Quoc V and Zhou, Denny and Chen, Xinyun},
  booktitle={The Twelfth International Conference on Learning Representations},
  year={2023}
}

@inproceedings{liang2024rich,
  title={Rich human feedback for text-to-image generation},
  author={Liang, Youwei and He, Junfeng and Li, Gang and Li, Peizhao and Klimovskiy, Arseniy and Carolan, Nicholas and Sun, Jiao and Pont-Tuset, Jordi and Young, Sarah and Yang, Feng and others},
  booktitle={Proceedings of the IEEE/CVF Conference on Computer Vision and Pattern Recognition},
  pages={19401--19411},
  year={2024}
}

@article{wang2023promptagent,
  title={Promptagent: Strategic planning with language models enables expert-level prompt optimization},
  author={Wang, Xinyuan and Li, Chenxi and Wang, Zhen and Bai, Fan and Luo, Haotian and Zhang, Jiayou and Jojic, Nebojsa and Xing, Eric P and Hu, Zhiting},
  journal={arXiv preprint arXiv:2310.16427},
  year={2023}
}

@article{yuksekgonul2024textgrad,
  title={Textgrad: Automatic" differentiation" via text},
  author={Yuksekgonul, Mert and Bianchi, Federico and Boen, Joseph and Liu, Sheng and Huang, Zhi and Guestrin, Carlos and Zou, James},
  journal={arXiv preprint arXiv:2406.07496},
  year={2024}
}

@misc{guo2025evopromptconnectingllmsevolutionary,
      title={EvoPrompt: Connecting LLMs with Evolutionary Algorithms Yields Powerful Prompt Optimizers}, 
      author={Qingyan Guo and Rui Wang and Junliang Guo and Bei Li and Kaitao Song and Xu Tan and Guoqing Liu and Jiang Bian and Yujiu Yang},
      year={2025},
      eprint={2309.08532},
      archivePrefix={arXiv},
      primaryClass={cs.CL},
      url={https://arxiv.org/abs/2309.08532}, 
}

@misc{cui2024phaseevounifiedincontextprompt,
      title={PhaseEvo: Towards Unified In-Context Prompt Optimization for Large Language Models}, 
      author={Wendi Cui and Jiaxin Zhang and Zhuohang Li and Hao Sun and Damien Lopez and Kamalika Das and Bradley Malin and Sricharan Kumar},
      year={2024},
      eprint={2402.11347},
      archivePrefix={arXiv},
      primaryClass={cs.CL},
      url={https://arxiv.org/abs/2402.11347}, 
}

@article{secheresse2025gaapo,
  title={GAAPO: Genetic Algorithmic Applied to Prompt Optimization},
  author={S{\'e}cheresse, Xavier and de Torcy, Antoine Villedieu and others},
  journal={arXiv preprint arXiv:2504.07157},
  year={2025}
}

@article{cho2023davidsonian,
  title={Davidsonian scene graph: Improving reliability in fine-grained evaluation for text-to-image generation},
  author={Cho, Jaemin and Hu, Yushi and Garg, Roopal and Anderson, Peter and Krishna, Ranjay and Baldridge, Jason and Bansal, Mohit and Pont-Tuset, Jordi and Wang, Su},
  journal={arXiv preprint arXiv:2310.18235},
  year={2023}
}

@article{huang2025t2i,
  title={T2i-compbench++: An enhanced and comprehensive benchmark for compositional text-to-image generation},
  author={Huang, Kaiyi and Duan, Chengqi and Sun, Kaiyue and Xie, Enze and Li, Zhenguo and Liu, Xihui},
  journal={IEEE Transactions on Pattern Analysis and Machine Intelligence},
  year={2025},
  publisher={IEEE}
}

@misc{schuhmann2022improved,
  author = {Schuhmann, Christoph},
  title = {{Improved Aesthetic Predictor}},
  howpublished = {\url{https://github.com/christophschuhmann/improved-aesthetic-predictor}},
  year = {2022},
  note = {GitHub repository},
}

@inproceedings{wu2023human,
  title={Human preference score: Better aligning text-to-image models with human preference},
  author={Wu, Xiaoshi and Sun, Keqiang and Zhu, Feng and Zhao, Rui and Li, Hongsheng},
  booktitle={Proceedings of the IEEE/CVF International Conference on Computer Vision},
  pages={2096--2105},
  year={2023}
}

@article{xu2023imagereward,
  title={Imagereward: Learning and evaluating human preferences for text-to-image generation},
  author={Xu, Jiazheng and Liu, Xiao and Wu, Yuchen and Tong, Yuxuan and Li, Qinkai and Ding, Ming and Tang, Jie and Dong, Yuxiao},
  journal={Advances in Neural Information Processing Systems},
  volume={36},
  pages={15903--15935},
  year={2023}
}

@inproceedings{chu2005preference,
  title={Preference learning with Gaussian processes},
  author={Chu, Wei and Ghahramani, Zoubin},
  booktitle={Proceedings of the 22nd international conference on Machine learning},
  pages={137--144},
  year={2005}
}

@article{eric2007active,
  title={Active preference learning with discrete choice data},
  author={Eric, Brochu and Freitas, Nando and Ghosh, Abhijeet},
  journal={Advances in neural information processing systems},
  volume={20},
  year={2007}
}

@article{koyama2020sequential,
  title={Sequential gallery for interactive visual design optimization},
  author={Koyama, Yuki and Sato, Issei and Goto, Masataka},
  journal={ACM Transactions on Graphics (TOG)},
  volume={39},
  number={4},
  pages={88--1},
  year={2020},
  publisher={ACM New York, NY, USA}
}

@article{koyama2017sequential,
  title={Sequential line search for efficient visual design optimization by crowds},
  author={Koyama, Yuki and Sato, Issei and Sakamoto, Daisuke and Igarashi, Takeo},
  journal={ACM Transactions on Graphics (TOG)},
  volume={36},
  number={4},
  pages={1--11},
  year={2017},
  publisher={ACM New York, NY, USA}
}

@article{li2025efficient,
  title={Efficient Visual Appearance Optimization by Learning from Prior Preferences},
  author={Li, Zhipeng and Liao, Yi-Chi and Holz, Christian},
  journal={arXiv preprint arXiv:2507.15355},
  year={2025}
}

@inproceedings{brochu2010bayesian,
  title={A Bayesian interactive optimization approach to procedural animation design},
  author={Brochu, Eric and Brochu, Tyson and De Freitas, Nando},
  booktitle={Proceedings of the 2010 ACM SIGGRAPH/Eurographics Symposium on Computer Animation},
  pages={103--112},
  year={2010}
}

@article{iwai2025constrained,
  title={Constrained Preferential Bayesian Optimization and Its Application in Banner Ad Design},
  author={Iwai, Koki and Kumagae, Yusuke and Koyama, Yuki and Hamasaki, Masahiro and Goto, Masataka},
  journal={arXiv preprint arXiv:2505.10954},
  year={2025}
}

@inproceedings{chan2022investigating,
  title={Investigating positive and negative qualities of human-in-the-loop optimization for designing interaction techniques},
  author={Chan, Liwei and Liao, Yi-Chi and Mo, George B and Dudley, John J and Cheng, Chun-Lien and Kristensson, Per Ola and Oulasvirta, Antti},
  booktitle={Proceedings of the 2022 CHI Conference on Human Factors in Computing Systems},
  pages={1--14},
  year={2022}
}

@article{yannakakis2015ratings,
  title={Ratings are overrated!},
  author={Yannakakis, Georgios N and Mart{\'\i}nez, H{\'e}ctor P},
  journal={Frontiers in ICT},
  volume={2},
  pages={13},
  year={2015},
  publisher={Frontiers Media SA}
}

@inproceedings{liao2024meta,
  title={A Meta-Bayesian Approach for Rapid Online Parametric Optimization for Wrist-based Interactions},
  author={Liao, Yi-Chi and Desai, Ruta and Pierce, Alec M and Taylor, Krista E and Benko, Hrvoje and Jonker, Tanya R and Gupta, Aakar},
  booktitle={Proceedings of the 2024 CHI Conference on Human Factors in Computing Systems},
  pages={1--38},
  year={2024}
}

@article{radford2019language,
  title={Language models are unsupervised multitask learners},
  author={Radford, Alec and Wu, Jeffrey and Child, Rewon and Luan, David and Amodei, Dario and Sutskever, Ilya and others},
  journal={OpenAI blog},
  volume={1},
  number={8},
  pages={9},
  year={2019}
}

@article{brown2020language,
  title={Language models are few-shot learners},
  author={Brown, Tom and Mann, Benjamin and Ryder, Nick and Subbiah, Melanie and Kaplan, Jared D and Dhariwal, Prafulla and Neelakantan, Arvind and Shyam, Pranav and Sastry, Girish and Askell, Amanda and others},
  journal={Advances in neural information processing systems},
  volume={33},
  pages={1877--1901},
  year={2020}
}

@article{fried2022incoder,
  title={Incoder: A generative model for code infilling and synthesis},
  author={Fried, Daniel and Aghajanyan, Armen and Lin, Jessy and Wang, Sida and Wallace, Eric and Shi, Freda and Zhong, Ruiqi and Yih, Wen-tau and Zettlemoyer, Luke and Lewis, Mike},
  journal={arXiv preprint arXiv:2204.05999},
  year={2022}
}

@article{jiang2024survey,
  title={A survey on large language models for code generation},
  author={Jiang, Juyong and Wang, Fan and Shen, Jiasi and Kim, Sungju and Kim, Sunghun},
  journal={arXiv preprint arXiv:2406.00515},
  year={2024}
}

@inproceedings{rombach2022high,
  title={High-resolution image synthesis with latent diffusion models},
  author={Rombach, Robin and Blattmann, Andreas and Lorenz, Dominik and Esser, Patrick and Ommer, Bj{\"o}rn},
  booktitle={Proceedings of the IEEE/CVF conference on computer vision and pattern recognition},
  pages={10684--10695},
  year={2022}
}

@article{dhariwal2021diffusion,
  title={Diffusion models beat gans on image synthesis},
  author={Dhariwal, Prafulla and Nichol, Alexander},
  journal={Advances in neural information processing systems},
  volume={34},
  pages={8780--8794},
  year={2021}
}

@article{singer2022make,
  title={Make-a-video: Text-to-video generation without text-video data},
  author={Singer, Uriel and Polyak, Adam and Hayes, Thomas and Yin, Xi and An, Jie and Zhang, Songyang and Hu, Qiyuan and Yang, Harry and Ashual, Oron and Gafni, Oran and others},
  journal={arXiv preprint arXiv:2209.14792},
  year={2022}
}

@article{ho2022imagen,
  title={Imagen video: High definition video generation with diffusion models},
  author={Ho, Jonathan and Chan, William and Saharia, Chitwan and Whang, Jay and Gao, Ruiqi and Gritsenko, Alexey and Kingma, Diederik P and Poole, Ben and Norouzi, Mohammad and Fleet, David J and others},
  journal={arXiv preprint arXiv:2210.02303},
  year={2022}
}

@article{sahoo2024systematic,
  title={A systematic survey of prompt engineering in large language models: Techniques and applications},
  author={Sahoo, Pranab and Singh, Ayush Kumar and Saha, Sriparna and Jain, Vinija and Mondal, Samrat and Chadha, Aman},
  journal={arXiv preprint arXiv:2402.07927},
  year={2024}
}

@inproceedings{DBLP:conf/chi/SubramonyamPPAS24,
  author       = {Hariharan Subramonyam and
                  Roy Pea and
                  Christopher Lawrence Pondoc and
                  Maneesh Agrawala and
                  Colleen M. Seifert},
  editor       = {Florian 'Floyd' Mueller and
                  Penny Kyburz and
                  Julie R. Williamson and
                  Corina Sas and
                  Max L. Wilson and
                  Phoebe O. Toups Dugas and
                  Irina Shklovski},
  title        = {Bridging the Gulf of Envisioning: Cognitive Challenges in Prompt Based
                  Interactions with LLMs},
  booktitle    = {Proceedings of the {CHI} Conference on Human Factors in Computing
                  Systems, {CHI} 2024, Honolulu, HI, USA, May 11-16, 2024},
  pages        = {1039:1--1039:19},
  publisher    = {{ACM}},
  year         = {2024},
  url          = {https://doi.org/10.1145/3613904.3642754},
  doi          = {10.1145/3613904.3642754},
  bibsource    = {dblp computer science bibliography, https://dblp.org}
}

@inproceedings{DBLP:conf/acl/QianHZDQCZZL0024,
  author       = {Cheng Qian and
                  Bingxiang He and
                  Zhong Zhuang and
                  Jia Deng and
                  Yujia Qin and
                  Xin Cong and
                  Zhong Zhang and
                  Jie Zhou and
                  Yankai Lin and
                  Zhiyuan Liu and
                  Maosong Sun},
  editor       = {Lun{-}Wei Ku and
                  Andre Martins and
                  Vivek Srikumar},
  title        = {Tell Me More! Towards Implicit User Intention Understanding of Language
                  Model Driven Agents},
  booktitle    = {Proceedings of the 62nd Annual Meeting of the Association for Computational
                  Linguistics (Volume 1: Long Papers), {ACL} 2024, Bangkok, Thailand,
                  August 11-16, 2024},
  pages        = {1088--1113},
  publisher    = {Association for Computational Linguistics},
  year         = {2024},
  url          = {https://doi.org/10.18653/v1/2024.acl-long.61},
  doi          = {10.18653/V1/2024.ACL-LONG.61},
  bibsource    = {dblp computer science bibliography, https://dblp.org}
}

@inproceedings{DBLP:conf/icml/RadfordKHRGASAM21,
  author       = {Alec Radford and
                  Jong Wook Kim and
                  Chris Hallacy and
                  Aditya Ramesh and
                  Gabriel Goh and
                  Sandhini Agarwal and
                  Girish Sastry and
                  Amanda Askell and
                  Pamela Mishkin and
                  Jack Clark and
                  Gretchen Krueger and
                  Ilya Sutskever},
  editor       = {Marina Meila and
                  Tong Zhang},
  title        = {Learning Transferable Visual Models From Natural Language Supervision},
  booktitle    = {Proceedings of the 38th International Conference on Machine Learning,
                  {ICML} 2021, 18-24 July 2021, Virtual Event},
  series       = {Proceedings of Machine Learning Research},
  volume       = {139},
  pages        = {8748--8763},
  publisher    = {{PMLR}},
  year         = {2021},
  url          = {http://proceedings.mlr.press/v139/radford21a.html},
  bibsource    = {dblp computer science bibliography, https://dblp.org}
}

@article{DBLP:journals/tochi/CherryL14,
  author       = {Erin Cherry and
                  Celine Latulipe},
  title        = {Quantifying the Creativity Support of Digital Tools through the Creativity
                  Support Index},
  journal      = {{ACM} Trans. Comput. Hum. Interact.},
  volume       = {21},
  number       = {4},
  pages        = {21:1--21:25},
  year         = {2014},
  url          = {https://doi.org/10.1145/2617588},
  doi          = {10.1145/2617588},
  bibsource    = {dblp computer science bibliography, https://dblp.org}
}

@incollection{HART1988139,
title = {Development of NASA-TLX (Task Load Index): Results of Empirical and Theoretical Research},
editor = {Peter A. Hancock and Najmedin Meshkati},
series = {Advances in Psychology},
publisher = {North-Holland},
volume = {52},
pages = {139-183},
year = {1988},
booktitle = {Human Mental Workload},
issn = {0166-4115},
doi = {https://doi.org/10.1016/S0166-4115(08)62386-9},
url = {https://www.sciencedirect.com/science/article/pii/S0166411508623869},
author = {Sandra G. Hart and Lowell E. Staveland}
}

@inproceedings{DBLP:conf/nips/ChristianoLBMLA17,
  author       = {Paul F. Christiano and
                  Jan Leike and
                  Tom B. Brown and
                  Miljan Martic and
                  Shane Legg and
                  Dario Amodei},
  editor       = {Isabelle Guyon and
                  Ulrike von Luxburg and
                  Samy Bengio and
                  Hanna M. Wallach and
                  Rob Fergus and
                  S. V. N. Vishwanathan and
                  Roman Garnett},
  title        = {Deep Reinforcement Learning from Human Preferences},
  booktitle    = {Advances in Neural Information Processing Systems 30: Annual Conference
                  on Neural Information Processing Systems 2017, December 4-9, 2017,
                  Long Beach, CA, {USA}},
  pages        = {4299--4307},
  year         = {2017},
  url          = {https://proceedings.neurips.cc/paper/2017/hash/d5e2c0adad503c91f91df240d0cd4e49-Abstract.html},
  bibsource    = {dblp computer science bibliography, https://dblp.org}
}

@inproceedings{nakashima2024swipeganspace,
  title={Swipeganspace: Swipe-to-compare image generation via efficient latent space exploration},
  author={Nakashima, Yuto and Yang, Mingzhe and Baba, Yukino},
  booktitle={Proceedings of the 29th International Conference on Intelligent User Interfaces},
  pages={675--685},
  year={2024}
}

@article{khattab2023dspy,
  title={Dspy: Compiling declarative language model calls into self-improving pipelines},
  author={Khattab, Omar and Singhvi, Arnav and Maheshwari, Paridhi and Zhang, Zhiyuan and Santhanam, Keshav and Vardhamanan, Sri and Haq, Saiful and Sharma, Ashutosh and Joshi, Thomas T and Moazam, Hanna and others},
  journal={arXiv preprint arXiv:2310.03714},
  year={2023}
}

@article{hahn2024proactive,
  title={Proactive agents for multi-turn text-to-image generation under uncertainty},
  author={Hahn, Meera and Zeng, Wenjun and Kannen, Nithish and Galt, Rich and Badola, Kartikeya and Kim, Been and Wang, Zi},
  journal={arXiv preprint arXiv:2412.06771},
  year={2024}
}

@article{cook2024ticking,
  title={Ticking all the boxes: Generated checklists improve llm evaluation and generation},
  author={Cook, Jonathan and Rockt{\"a}schel, Tim and Foerster, Jakob and Aumiller, Dennis and Wang, Alex},
  journal={arXiv preprint arXiv:2410.03608},
  year={2024}
}

@article{viswanathan2025checklists,
  title={Checklists are better than reward models for aligning language models},
  author={Viswanathan, Vijay and Sun, Yanchao and Ma, Shuang and Kong, Xiang and Cao, Meng and Neubig, Graham and Wu, Tongshuang},
  journal={arXiv preprint arXiv:2507.18624},
  year={2025}
}

@article{gao2024aligning,
  title={Aligning llm agents by learning latent preference from user edits},
  author={Gao, Ge and Taymanov, Alexey and Salinas, Eduardo and Mineiro, Paul and Misra, Dipendra},
  journal={Advances in Neural Information Processing Systems},
  volume={37},
  pages={136873--136896},
  year={2024}
}

@inproceedings{petridis2024constitutionmaker,
  title={Constitutionmaker: Interactively critiquing large language models by converting feedback into principles},
  author={Petridis, Savvas and Wedin, Benjamin D and Wexler, James and Pushkarna, Mahima and Donsbach, Aaron and Goyal, Nitesh and Cai, Carrie J and Terry, Michael},
  booktitle={Proceedings of the 29th International Conference on Intelligent User Interfaces},
  pages={853--868},
  year={2024}
}

@article{wang2025promptimizer,
  title={Promptimizer: User-Led Prompt Optimization for Personal Content Classification},
  author={Wang, Leijie and Yurechko, Kathryn and Zhang, Amy X},
  journal={arXiv preprint arXiv:2510.09009},
  year={2025}
}

@article{huang2018music,
  title={Music transformer},
  author={Huang, Cheng-Zhi Anna and Vaswani, Ashish and Uszkoreit, Jakob and Shazeer, Noam and Simon, Ian and Hawthorne, Curtis and Dai, Andrew M and Hoffman, Matthew D and Dinculescu, Monica and Eck, Douglas},
  journal={arXiv preprint arXiv:1809.04281},
  year={2018}
}

@inproceedings{yao2022react,
  title={React: Synergizing reasoning and acting in language models},
  author={Yao, Shunyu and Zhao, Jeffrey and Yu, Dian and Du, Nan and Shafran, Izhak and Narasimhan, Karthik R and Cao, Yuan},
  booktitle={The eleventh international conference on learning representations},
  year={2022}
}

@article{wei2022chain,
  title={Chain-of-thought prompting elicits reasoning in large language models},
  author={Wei, Jason and Wang, Xuezhi and Schuurmans, Dale and Bosma, Maarten and Xia, Fei and Chi, Ed and Le, Quoc V and Zhou, Denny and others},
  journal={Advances in neural information processing systems},
  volume={35},
  pages={24824--24837},
  year={2022}
}
